\documentclass[lettersize, journal]{IEEEtran}

\usepackage{float}
\usepackage{caption}
\usepackage{soul, color, xcolor}
\usepackage{hyperref}
\hypersetup{hidelinks}

\usepackage{cite}
\usepackage{amsmath,amssymb,amsfonts}
\expandafter\let\csname equation*\endcsname\relax
\expandafter\let\csname endequation*\endcsname\relax
\usepackage{amsthm}
\usepackage{graphicx}
\usepackage{float} 
\usepackage{textcomp}
\usepackage{svg}
\usepackage{algorithm}
\usepackage{mathtools}

\usepackage{algpseudocode}
\usepackage{subcaption}
\newtheorem{theorem}{Theorem}

\newtheorem{lemma}{Lemma}

\newtheorem{corollary}{Corollary}

\allowdisplaybreaks
\makeatletter
\def\ScaleIfNeeded{
\ifdim\Gin@nat@ width>\linewidth \linewidth \else \Gin@nat@width
\fi } \makeatother

\graphicspath{{Fig/}}

\begin{document}

\title{Movable-Element STARS-Aided Secure Communications}

\author{Jingjing Zhao, Qian Xu, Kaiquan Cai, Yanbo Zhu, Xidong Mu, and Yuanwei Liu,~\IEEEmembership{Fellow,~IEEE}
\thanks{J. Zhao, Q. Xu, K. Cai, and Y. Zhu are with the School of Electronic and Information Engineering and the State Key Laborotary of CNS/ATM, Beihang University, Beijing, China (email:\{jingjingzhao, 15266317562, caikq, zhuyanbo\}@buaa.edu.cn).} 
\thanks{X. Mu is with the Centre for Wireless Innovation (CWI), Queen's University Belfast, Belfast, BT3 9DT, U.K. (e-mail: x.mu@qub.ac.uk).}
\thanks{Y. Liu is with the Department of Electrical and Electronic Engineering, the University of Hong Kong, Hong Kong, China (e-mail: yuanwei@hku.hk).}
}

\maketitle

\begin{abstract}
A novel movable-element (ME) enabled simultaneously transmitting and reflecting surface (ME-STARS)-aided secure communication system is investigated. Against the full-space eavesdropping, MEs are deployed at the STARS for enhancing the physical layer security by exploiting higher spatial degrees of freedom. Specifically, a sum secrecy rate maximization problem is formulated, which jointly optimizes the passive beamforming and the MEs positions at the ME-STARS, as well as the active beamforming at the base station. To solve the resultant non-convex optimization problem involving highly-coupled variables, an alternating optimization-based iterative algorithm is developed, decomposing the original problem into three subproblems. In particular, for the MEs position optimization subproblem, a gradient ascent algorithm is employed to iteratively refine the MEs' locations within the confined region. Moreover, the the active and passive beamforming subproblems are solved by employing successive convex approximation. Numerical results unveil that: 1) ME-STARS significantly improves the secrecy performance compared to the conventional STARS with fixed-position elements; and 2) The secrecy rate achieved by the ME-STARS gets saturated within limited movable region size.

\end{abstract}


\section{Introduction}

Reconfigurable intelligent surfaces (RISs) have garnered significant attention in recent years as an promising technology in wireless communications~\cite{zeng2020reconfigurable}. Compared to conventional multiple-input multiple-output (MIMO) systems, RISs can dynamically adjust the incident signal by customizing the phase response of each element with the aid of a smart controller~\cite{di2022communication}. However, reflection/transmission-only RISs~\cite{tang2020wireless} can only provide half-space coverage, which restricts the flexibility and effectiveness for the practical implementation. To overcome this limitation, the concept of simultaneously transmitting and reflecting surface (STARS) has been proposed~\cite{liu2021star, mu2024simultaneously, xu2021star}. Different form traditional RISs, STARS enhances the communication performance by splitting the incident signal into both transmitted and reflected signals on each side of the surface, thereby providing a full-space communication environment. Consequently, STARS offers additional degrees of freedom (DoFs) in controlling signal propagation, which in turn improves the flexibility in wireless communication networks~\cite{mu2021simultaneously}.

The conventional RISs and STARS usually deploy fixed-position antennas (FPAs), which limits their ability to exploit channel variations in the spatial domain. {To further exploit the spatial diversity}, movable antenna (MA) technology has emerged as a promising solution~\cite{zhu2023modeling, ma2023mimo}.  Unlike FPAs, each MA is connected to the radio frequency (RF) chain via a flexible cable, allowing it to be repositioned within a range of several to tens of wavelengths, so as to dynamically adapt to the time-varying propagation environment~\cite{zhu2023movableopti}. The MA technique can effectively avoid deep-fading effect and reconstruct more favorable channel conditions. Therefore, deploying MAs at the RIS/STARS can play a crucial role in enhancing the system performance and the flexibility.

Due to broadcast nature of wireless communications, secure transmission becomes a critical issue. Physical layer security (PLS) emerges as a key technology to address this issue in wireless communications~\cite{zou2016survey, mukherjee2014principles, liu2016physical}. Particularly, given the STARS capability of manipulating both the transmission and reflection signals, potential eavesdroppers can intercept information from both sides of the STARS, which leads to the phenomenon of \textit{full-space eavesdropping}. As such, the security issue becomes more stringent and complicated, which requires more advanced PLS techniques for securing the STARS-aided communications. Fortunately, with the aid of MA techniques, the enhanced spatial DoFs can help to enhance the communication quality for legitimate users, while weakening the signals received by eavesdroppers~\cite{zhang2023security}. By efficiently exploiting the channel variations in the continuous spatial domain through MAs, new opportunities are opened for secure communication in STARS-aided systems, enhancing protection against eavesdropping threats.

\subsection{Related Works}
\subsubsection{Studies on RIS-aided PLS} 
In recent studies, the application of RIS for enhancing PLS has received significant attention and been extensively explored. In~\cite{cui2019secure}, the authors investigated how jointly optimizing both RIS reflective beamforming and base station (BS) transmit beamforming could maximize secrecy rates, particularly in scenarios where the channels of legitimate users and eavesdroppers were highly correlated. Building on this, the authors of~\cite{yu2020robust} developed a robust beamforming strategy for multi-RIS-assisted multiuser systems, addressing the challenges posed by imperfect channel state information (CSI). A RIS-assisted system was investigated in~\cite{guan2020intelligent}, where an artificial noise (AN)-aided secure transmission scheme demonstrated the effectiveness of AN in mitigating eavesdropping risks. Additionally, the authors of~\cite{liu2021detect} proposed a three-step training mechanism to detect pilot spoofing attacks and estimate cascaded CSI, although their approach did not incorporate direct channel estimation or secure beamforming design. Extending these efforts, the authors of \cite{zhang2020robust} explored a RIS-assisted non-orthogonal multiple access (NOMA) network and highlighted the benefits of the beamforming scheme using AN for secure transmission. The authors of \cite{zhang2021securing} designed a robust secure beamforming framework for RIS-aided NOMA networks, emphasizing the importance of optimizing RIS configurations to counteract passive eavesdropping. A stochastic geometry-based analytical framework for secure STARS-assisted NOMA transmissions was proposed in~\cite{xie2023physical}, which considered both time-switching and energy-splitting protocols, providing closed-form expressions for secrecy outage probability and average secrecy capacity. The utilization of STARS for joint PLS and covert communications in a multi-antenna millimeterwave system was investigated in~\cite{xiao2024star}, optimizing covert and secrecy rates while ensuring quality of service (QoS) for legal users. A STARS-aided physical-layer key generation framework was proposed in~\cite{song2024phase}, where phase-shift adjustments was used to minimize the average bit disagreement ratio. 
The STARS-assisted NOMA system with cooperative jammer and dual eavesdroppers was studied in~\cite{qin2024deep}, where a deep deterministic policy gradient-based algorithm was proposed to maximize the sum secrecy rate under perfect/imperfect CSI.

\subsubsection{Studies on MA communications}
The MA position design offers several advantages over conventional FPA systems, including improved signal power, better interference mitigation, and enhanced flexible beamforming~\cite{zhu2023movable}. The authors first explored the channel model for MA-aided systems in~\cite{zhu2023modeling}, proposing a field-based channel model applicable to far-field scenarios. 
By analyzing performance across deterministic and statistical channels, they showed that MA systems achieved superior results compared to FPA systems. The authors of~\cite{ma2024multi} demonstrated the superior performance of MA-assisted systems in enhancing multi-beamforming. In \cite{hu2024movable}, the authors explored a MA-enabled wireless communication system that enhances coordinated multipoint transmission by maximizing the effective received signal-to-noise ratio. Moreover, in \cite{zhu2024performance}, the authors demostrated that the MA-aided system boosted broadband communication performance by adjusting the ME positions, thereby providing additional DoFs.
Building on these advancements, recent research has focused on the integration of MA with RIS, which can bring greater performance improvements. The integration was investigated in~\cite{sun2024sum}, where fractional programming technique was developed to optimize the positions of MAs, the beamforming matrix of BS, and the reflection coefficients of RIS for maximizing the sum-rate. Additionally, the authors of~\cite{zhang2024sum} proposed an active RIS-assisted MA system, demonstrating its ability to significantly enhance multi-user communication by maximizing the sum-rate in complex scattering environments, which outperformed the fixed position element scheme. In~\cite{xie2024movable}, a covert communication system assisted by both MA and RIS was studied, where deep reinforcement learning was used to alternatively optimize the system's subproblems and maximize the covert rate. Moreover,~\cite{zhao2024exploiting} explored the integration of STARS and MA, demonstrating that their combined use significantly improves the system's weighted sum-rate through joint optimization under the three operating protocols - energy splitting, mode switching, ans time switching.

\subsection{Motivations and Contributions}
The unique feature of the STARS enabled full-space propagation leads to the potential of mutual eavesdropping, i.e., each eavesdropper is capable of intercepting signals intended for users on both sides of the STARS, which introduces more challenging secrecy issues. To address this issue, the movable element (ME)-STARS provides higher spatial-domain DoFs by allowing MEs to move to preferable positions, so as to enhance the intended signals strength and suppress the information leakage. However, since the MEs positions affect the cascaded channels between the BS, the STARS, and legitimate users/eavesdroppers, which makes the MEs positions optimization more challenging. To the best of our knowledge, the application of ME-STARS in PLS systems has not been investigated yet, which provides the main motivation for this work. 

This paper proposes a novel ME-STARS-aided secure communication framework, where a ME-STARS is employed for desired signal enhancement and anti-eavesdropping in a full-space manner. 
We investigate the joint MEs positions and active/passive beamforming optimization problem with the aim of maximizing the sum secrecy rate. The main contributions are outlined below.

\begin{itemize}
    \item We propose a ME-STARS-aided secure communications framework, where MEs positions can be flexibly adjusted to enhance the spatial DoFs under the presence of multiple eavesdroppers. Based on this framework, a joint optimization problem of MEs positions and active/passive beamforming is formulated to maximize the sum secrecy rate, subject to minimum rate requirements of legitimate users and maximum leakage rate constraints at eavesdroppers.                              
    \item We propose an alternating optimization (AO) method for iteratively updating the involved variables of the resultant non-convex coupled optimization problem. We invoke the gradient ascent algorithm combined with the penalty method for addressing the MEs positions optimization subproblem. 
    To address the active and passive beamforming subproblems, we employ the successive convex approximation (SCA) technique. Moreover, we provide a convergence analysis to demonstrate that the proposed AO method converges to a stable solution.
    \item  Numerical results verify that: 1) the proposed ME-STARS scheme significantly outperforms the conventional FPE-STARS as well as the ME-RIS in terms of the secure communications performance; and 2) the sum secrecy rate achieved by the ME-STARS increases with the enlarged movable region size and gradually saturates within a limited value.

\end{itemize}

\subsection{Organization and Notation}
The remainder of this paper is outlined as follows. Section II presents the ME-STARS-aided secure communication system model with field-response channels, and formulates the problem of maximizing the sum secrecy rate. Section III develops an AO-based iterative algorithm for jointly optimizing MEs positions and active/passive beamforming. Section IV presents simulation results to demonstrate the effectiveness of the proposed framework and algorithms. Finally, Section V concludes the paper. 

\textit{Notations}: Scalars are represented by italicized characters, while bold letters are used for vectors and matrices, with lowercase letters for vectors and uppercase ones for matrices. The space of $N \times M$ complex-valued matrices is indicated by $\mathbb{C}^{N \times M}$. For a vector $\mathbf{x}$, $\mathbf{x}^*$ signifies the conjugate of $\mathbf{x}$, $\mathbf{x}^H$ represents the Hermitian (conjugate) transpose, and $\|\mathbf{x}\|$ its Euclidean norm. The term $\text{diag}(\mathbf{x})$ refers to a diagonal matrix with the elements of $\mathbf{x}$on its main diagonal. A circularly symmetric complex Gaussian (CSCG) random variable with mean $\mu$ and variance $\sigma^2$ is expressed as $\mathcal{C}\mathcal{N}(\mu, \sigma^2)$. The set of all $N$-dimensional complex Hermitian matrices is denoted by $\mathbb{H}^N$. For a matrix $\mathbf{A}$, $\text{Rank}(\mathbf{A})$ represents its rank, $\text{Tr}(\mathbf{A})$ its trace, and $\text{Diag}(\mathbf{A})$ a vector comprising its main diagonal elements. The notation $\mathbf{A} \succeq 0$ signifies that $\mathbf{A}$ is positive semidefinite. The nuclear, spectral, and Frobenius norms of $\mathbf{A}$ are represented by $\|\mathbf{A}\|_*$, $\|\mathbf{A}\|_2$, and $\|\mathbf{A}\|_F$, respectively. A vector of size $M \times 1$ where all elements are equal to 1 is denoted as $\mathbf{1}_{M \times 1}$. Finally, the symbol $\nabla$ indicates the gradient, and  the symbol $\circ$ denotes the Hadamard product.

\section{System Model and Problem Formulation}
In this section, we first introduce the ME-STARS-aided secure communication system model with field-response channels. Subsequently, we formulate the joint MEs positions and active/passive beamforming optimization problem aimed at maximizing the sum secrecy rate. 

\subsection{System Model}

\begin{figure}[h]
    \centering
    \includegraphics[scale=0.27]{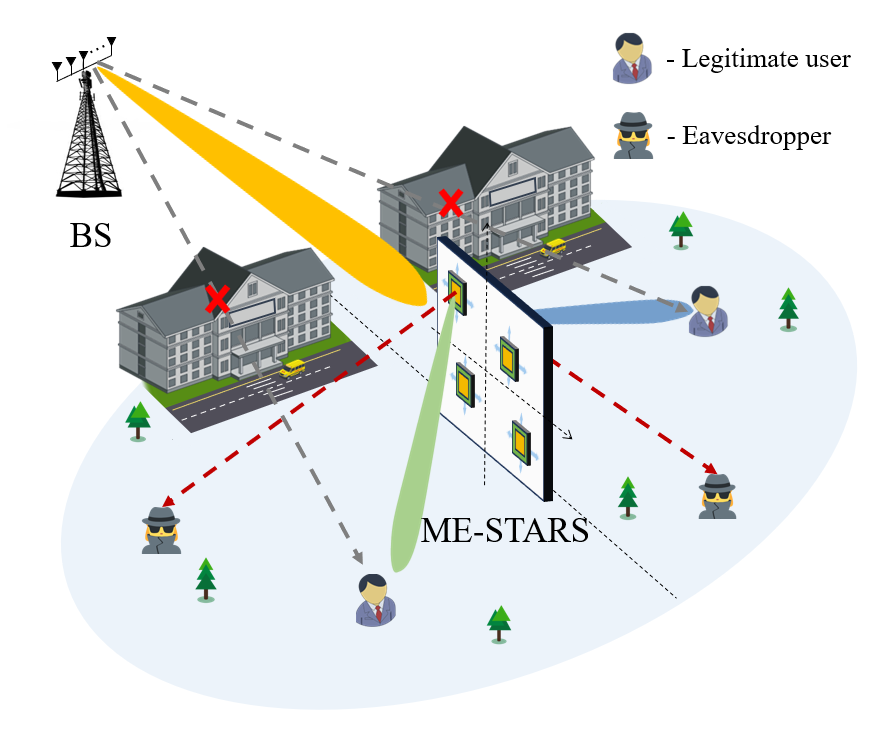}
    \caption{Illustration of the ME-STARS-aided downlink MISO communication system.}
    \label{fig: system model}
\end{figure}
As illustrated in Fig.~\ref{fig: system model}, we consider a ME-STARS-aided downlink MISO communication system, which consists of a BS equipped with $M$ FPAs, a STARS equipped with $N$ MEs, two single antenna legitimate users, and two single antenna eavesdroppers. For the ease of presentation and revealing the fundamental insights, we consider the setup of one legitimate user $l_{\text{t}}$ and one eavesdropper $e_{\text{t}}$ located in the transmission space, as well as one legitimate user $l_{\text{r}}$ and one eavesdropper $e_{\text{r}}$ located in the reflection space. Let $\mathcal{K}_{\text{l}} = \{l_{\text{t}}, l_{\text{r}}\}$ and $\mathcal{K}_{\text{e}} = \{e_{\text{t}}, e_{\text{r}}\}$ represent the set of legitimate users and eavesdroppers, respectively, and $\mathcal{K} = \mathcal{K}_{\text{l}} \cup \mathcal{K}_{\text{e}}$.
The STARS transmission and reflection coefficient matrices are denoted by $\mathbf{\Theta}_{\text{t}}=\text{diag}\left(\sqrt{\beta_1^{\text{t}}}e^{j\theta_1^{\text{t}}}, \sqrt{\beta_2^{\text{t}}}e^{j\theta_2^{\text{t}}},\ldots, \sqrt{\beta_N^{\text{t}}}e^{j\theta_N^{\text{t}}} \right)$ and $\mathbf{\Theta}_{\text{r}}=\text{diag}\left(\sqrt{\beta_1^{\text{r}}}e^{j\theta_1^{\text{r}}},\sqrt{\beta_2^{\text{r}}}e^{j\theta_2^{\text{r}}},\ldots, \sqrt{\beta_N^{\text{r}}}e^{j\theta_N^{\text{r}}}\right)$, respectively, where $\beta_n^{\text{t}}, \beta_n^{\text{r}}\in [0, 1]$ represent energy splitting coefficients, satisfying $\beta_n^{\text{t}}+\beta_n^{\text{r}}=1$, and $\theta_n^{\text{t}}$, $\theta_n^{\text{r}}\in [0, 2\pi)$ represent phase shifts for the transmission and reflection signals, respectively. The BS FPAs and STARS MEs are located in the $x_b - O_b - y_b$ and $x_s - O_s - y_s$ planes, with their local original points denoted by $O_b=[0, 0]^T$ and $O_s=[0, 0]^T$, respectively. The local coordinates of the $m$-th FPA at the BS and the $n$-th ME at the STARS are denoted by $\mathbf{t}_m=\left[x_m,y_m\right]^T$, and $\mathbf{r}_n=\left[x_n,y_n\right]^T\in\mathcal{C}$, respectively, where $\mathcal{C}$ denotes the movable region.  Further define $\mathbf{R}=\left[\mathbf{r}_1, \mathbf{r}_2, \ldots, \mathbf{r}_N\right]\in\mathbb{R}^{2\times N}$ as the STARS element position matrix (EPM). Leveraging MEs at the STARS, the communication environment between the BS and legitimate users/eavesdroppers can be reconfigured not only by the transmission and reflection coefficients, but also by the adjustable channel response achieved with flexible MEs positions.

\subsubsection{Channel and Signal Model}

\begin{figure}[ht]
    \centering
    \includegraphics[scale=0.15]{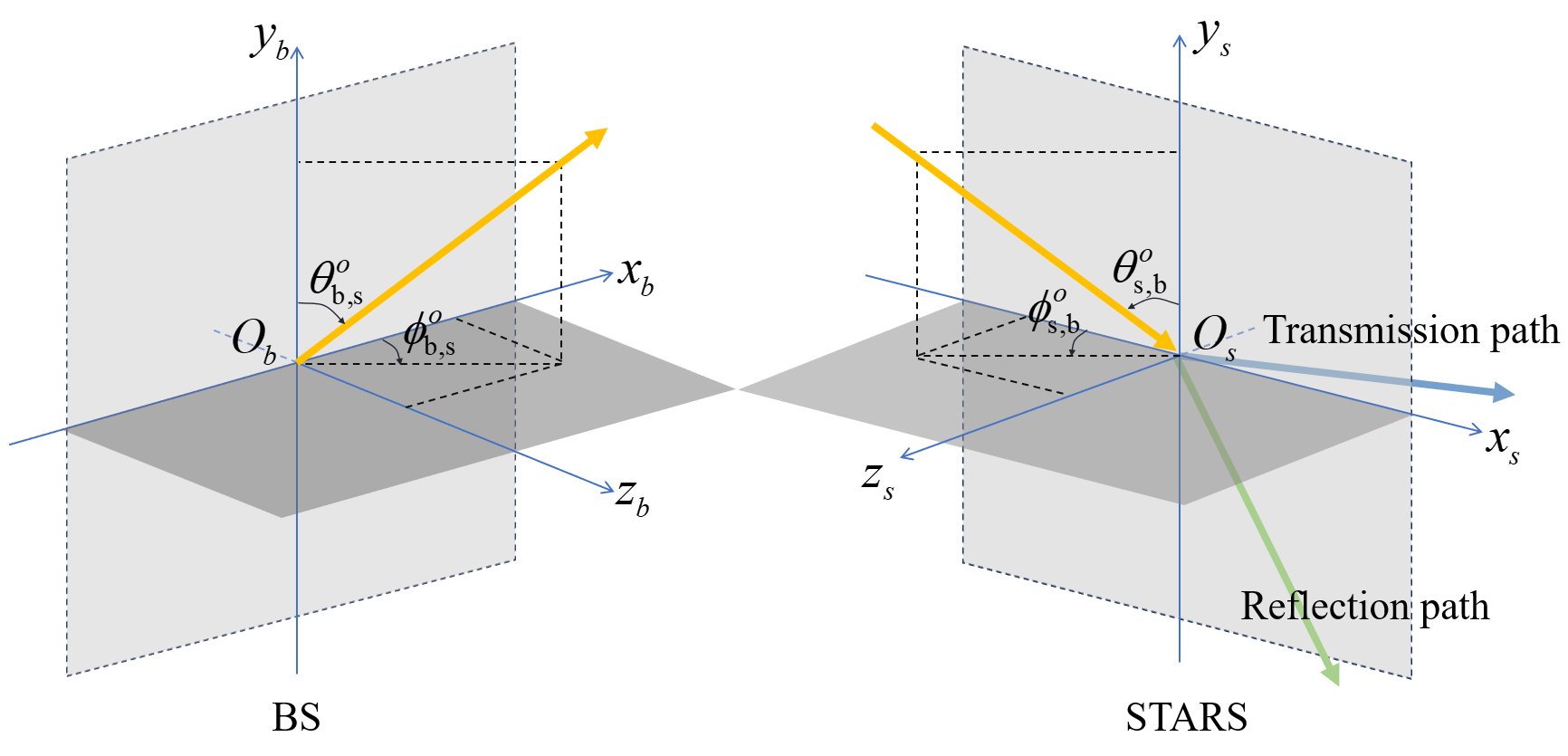}
    \caption{The spatial angles at the BS and the STARS.}
    \label{fig: The far-field-response channel model}
\end{figure}

We assume that obstacles block the direct communication links between the BS and legitimate users/eavesdroppers. Let $L_{\text{BS}}$ denotes the number of paths from the BS to the STARS, and $L_{\text{S}, k}, k \in \mathcal{K}$ represents the number of paths from the STARS to legitimate users/eavesdroppers. 
Consequently, as shown in Fig.~\ref{fig: The far-field-response channel model}, the field response vector (FRV)~\cite{ma2023mimo} of the $m$-th antenna at the BS is given by
\begin{equation}
\label{eq: FRV of the m-th antenna at the BS}
    \mathbf{e}(\mathbf{t}_m) = \left[e^{j\frac{2\pi}{\lambda}\rho_{\text{B}}^1(\mathbf{t}_m)}, e^{j\frac{2\pi}{\lambda}\rho_{\text{B}}^2(\mathbf{t}_m)}, \ldots, e^{j\frac{2\pi}{\lambda}\rho_{\text{B}}^{L_{\text{BS}}}(\mathbf{t}_m)}\right]^T,
\end{equation}
where $\rho_{\text{B}}^o(\mathbf{t}_m) = x_m\text{cos}\theta_{\text{b,s}}^o\text{sin}\phi_{\text{b,s}}^o + y_m\text{sin}\theta_{\text{b,s}}^o$ is the signal propagation difference for the $o$-th path between the reference point $O_{\text{b}}$ and the $m$-th FPA position $\mathbf{t}_m$, with $\theta_{\text{b,s}}^o$ and $\phi_{\text{b,s}}^o$ denoting the elevation and azimuth angle of departure (AoD) for the $o$-th path between the BS and the STARS, respectively. Accordingly, the field response matrix for all FPAs at the BS can be represented by $\mathbf{E} = \left[\mathbf{e}(\mathbf{t}_1), \mathbf{e}(\mathbf{t}_2), \ldots, \mathbf{e}(\mathbf{t}_M)\right]\in\mathbb{C}^{L_{\text{BS}} \times M}$, which is a constant matrix due to the fixed positions of the antennas at the BS. 

Under the plane-wave propagation assumption, when MEs move within the given STARS region, angles and channel gain amplitudes remain constant across all paths, while only the channel gain phases vary. Therefore, the FRV for the $n$-th ME at the STARS, corresponding to the incident channel, is given by
\begin{equation}
\label{eq: FRV for the n-th ME at the STARS}
    \mathbf{f}_{\text{in}}(\mathbf{r}_n) = \left[e^{j\frac{2\pi}{\lambda}\rho_{\text{S,in}}^1(\mathbf{r}_n)}, e^{j\frac{2\pi}{\lambda}\rho_{\text{S,in}}^2(\mathbf{r}_n)}, \ldots, e^{j\frac{2\pi}{\lambda}\rho_{\text{S,in}}^{L_{\text{BS}}}(\mathbf{r}_n)}\right]^T,
\end{equation}
where $\rho_{\text{S,in}}^o(\mathbf{r}_n) = x_n\text{cos}\theta_{\text{s,b}}^o\text{sin}\phi_{\text{s,b}}^o + y_n\text{sin}\theta_{\text{s,b}}^o$ represents the signal propagation difference for the $o$-th path between the reference point $O_s$ an the $n$-th ME position $\mathbf{r}_n$, with $\theta_{\text{s,b}}^o$ and $\phi_{\text{s,b}}^o$ denoting the elevation and  azimuth angle of arrival (AoA) of the $o$-th path between the STARS and the BS, respectively. By aggregating all the FRVs, the field response matrix at the STARS for the incident channel can be expressed as $\mathbf{F}_{\text{in}}(\mathbf{R}) = \left[\mathbf{f}_{\text{in}}(\mathbf{r}_1), \mathbf{f}_{\text{in}}(\mathbf{r}_2), \ldots, \mathbf{f}_{\text{in}}(\mathbf{r}_N)\right]\in\mathbb{C}^{L_{\text{BS}} \times N}$. Let $\boldsymbol{\Sigma}_{\text{BS}}\in\mathbb{C}^{L_{\text{BS}} \times L_{\text{BS}}}$ denote the response from the reference point $O_{\text{b}}$ at the BS to reference point $O_s$ at the STARS. Then, the channel response from the BS to the STARS can be represented as
\begin{equation}
\label{eq: channel response from the BS to the STARS}
    \mathbf{H}(\mathbf{R}) = \left(\mathbf{F}_{\text{in}}(\mathbf{R})\right)^H\boldsymbol{\Sigma}_{\text{BS}}\mathbf{E}.
\end{equation}

Similarly, the FRV of the $n$-th ME at the STARS for the legitimate users/eavesdroppers can be expressed as 
\begin{equation}
\label{eq: FRV of the n-th ME at the STARS for the legitimate user}
    \mathbf{f}_{k}(\mathbf{r}_n) = \left[e^{j\frac{2\pi}{\lambda}\rho_{\text{S}, k}^1(\mathbf{r}_n)}, e^{j\frac{2\pi}{\lambda}\rho_{\text{S}, k}^2(\mathbf{r}_n)}, \ldots, e^{j\frac{2\pi}{\lambda}\rho_{\text{S}, k}^{L_{\text{S},k}}(\mathbf{r}_n)}\right]^T,
\end{equation}
where $k\in\mathcal{K}$, $\rho_{\text{S},k}^p(\mathbf{r}_n) = x_n\text{cos}\theta_{\text{S}, k}^p\text{sin}\phi_{\text{S}, k}^p + y_n\text{sin}\theta_{\text{S}, k}^p$ is the signal propagation difference, with $\theta_{\text{S}, k}^p$ and $\phi_{\text{S}, k}^p$ denoting the elevation and azimuth AoDs for the $p$-th path between the STARS and the legitimate users/eavesdroppers, respectively. 
Then, we define $\sigma_{\text{S}, k}\in\mathbb{C}^{1\times L_{\text{S},k}}$ as the response from the STARS reference point $O_s$ to the legitimate users/eavesdroppers. We obtain the STARS to legitimate users/eavesdroppers channel response as 
\begin{equation}
\label{eq:  STARS to legitimate user channel response}
    \mathbf{h}_{k_{\text{l}}}(\mathbf{R}) = \sigma_{\text{S}, k_{\text{l}}}\mathbf{F}_{k_{\text{l}}}(\mathbf{R}), \forall k_{\text{l}}\in\mathcal{K}_{\text{l}},
\end{equation}

\begin{equation}
\label{eq:  STARS to Eve channel response}
    \mathbf{g}_{k_\text{e}}(\mathbf{R}) = \sigma_{\text{S}, k_{\text{e}}}\mathbf{F}_{k_{\text{e}}}(\mathbf{R}), \forall k_{\text{e}}\in\mathcal{K}_{\text{e}},
\end{equation}
where $\mathbf{F}_{k}(\mathbf{R}) = \left[\mathbf{f}_{k}(\mathbf{r}_1), \mathbf{f}_{k}(\mathbf{r}_2), \ldots, \mathbf{f}_{k}(\mathbf{r}_N)\right]$, $\forall k\in\mathcal{K}$.

\subsubsection{Communication Model}

Based on the above signal and channel models, the signal received by the legitimate users and eavesdroppers can be expressed by 
\begin{equation}
\label{eq: the signal received by the legitimate user}
    y^{\text{l}}_{k_{\text{l}}} = \mathbf{h}_{k_{\text{l}}}(\mathbf{R})\mathbf{\Theta}_{\varrho}\mathbf{H}(\mathbf{R})\left( \sum_{k_{\text{l}}\in\mathcal{K}_{\text{l}}}\mathbf{w}_{k_{\text{l}}}x_{k_{\text{l}}} \right) + n_{{\text{l}}},
\end{equation}
and
\begin{equation}
\label{eq: the signal received by the Eve}
    y^{\text{e}}_{k_{\text{e}},k_{\text{l}}} = \mathbf{g}_{k_{\text{e}}}(\mathbf{R})\mathbf{\Theta}_{\varrho}\mathbf{H}(\mathbf{R})\left( \sum_{k_{\text{l}}\in\mathcal{K}_{\text{l}}}\mathbf{w}_{k_{\text{l}}}x_{k_{\text{l}}} \right) + n_{{\text{e}}},
\end{equation}
respectively, where $\varrho \in \{\text{t}, \text{r}\}$, $\mathbf{w}_{k_{\text{l}}}\in\mathbb{C}^{M\times1}$ represents the BS beamforming vector for the signal intended to the legitimate user $k_{\text{l}}$, $x_{k_{\text{l}}}$ denotes the transmission signal to the legitimate user $k_{\text{l}}$, $n_{{\text{l}}}\sim\mathcal{CN}\left(0,\sigma_{\text{l}}^2\right)$ and $n_{{\text{e}}}\sim\mathcal{CN}\left(0,\sigma_{\text{e}}^2\right)$ are the zero-mean additive white Gaussian noise (AWGN) at the legitimate users and eavesdroppers, respectively. Accordingly, the achievable data rate at the legitimate user $k_{\text{l}}$ is given by
\begin{equation}
\label{eq: legitimate users achievable rate}
    R_{k_{\text{l}}} = \text{log}_2\left(1+\frac{\left|\mathbf{h}_{k_{\text{l}}}(\mathbf{R})\mathbf{\Theta}_{\varrho}\mathbf{H}(\mathbf{R})\mathbf{w}_{k_{\text{l}}}\right|^2}{\left|\mathbf{h}_{k_{\text{l}}}(\mathbf{R})\mathbf{\Theta}_{\varrho}\mathbf{H}(\mathbf{R})\mathbf{w}_{\bar{k}_{\text{l}}}\right|^2 + \sigma_{\text{l}}^2}\right).
\end{equation}
Moreover, the data rate for the eavesdropper $k_{\text{e}}$ for wiretapping the information signal of $k_{\text{l}}$ is given by
\begin{equation}
\label{eq: Eves achievable rate}
     R^{\text{e}}_{k_{\text{e}},k_{\text{l}}} = \text{log}_2\left(1+\frac{\left|\mathbf{g}_{k_{\text{e}}}(\mathbf{R})\mathbf{\Theta}_{\varrho}\mathbf{H}(\mathbf{R})\mathbf{w}_{k_{\text{l}}}\right|^2}{\left|\mathbf{g}_{k_{\text{e}}}(\mathbf{R})\mathbf{\Theta}_{\varrho}\mathbf{H}(\mathbf{R})\mathbf{w}_{\bar{k}_{\text{l}}}\right|^2 + \sigma_{\text{e}}^2}\right).
\end{equation}
Then, we define the secrecy communication rate of the legitimate user $k_{\text{l}}$ as
\begin{equation}
\label{eq: secrecy communication rate}
    R^{\text{s}}_{k_{\text{l}}} = \left[R_{k_{\text{l}}} - \text{max}\left\{R^{\text{e}}_{e_{\text{t}}, k_{\text{l}}}, R^{\text{e}}_{e_{\text{r}}, k_{\text{l}}}\right\}\right]^+, \forall k_{\text{l}}\in\mathcal{K}_{\text{l}},
\end{equation}
where $[x]^+ = \text{max}\{x,0\}$.

\subsection{Problem Formulation}
In this paper, we aim to maximize the sum secrecy rate by jointly optimizing the MEs positions and the passive beamforming at the STARS, and the active beamforming at the BS. Let $\mathbf{W} = \left[\mathbf{w}_{l_{\text{t}}}, \mathbf{w}_{l_{\text{r}}}\right] \in \mathbb{C}^{M \times 2}$ represent the beamforming matrix of the BS. The sum secrecy rate maximization problem can be formulated as
\begin{subequations}
\label{eq: optimization problem}
\begin{equation}
\label{eq: secrecy rate_1}
    \max_{\left\{\mathbf{R}, \boldsymbol{\Theta}_{\varrho}, \mathbf{W} \right\}} \sum_{k_{\text{l}}\in\mathcal{K}_{\text{l}}} R^{\text{s}}_{k_{\text{l}}} 
\end{equation}
\begin{equation}
\label{eq: phase constraints}
    {\rm{s.t.}} \ \  \theta_n^{\text{t}}, \theta_n^{\text{r}} \in [0, 2\pi), 1\leq n\leq N,
\end{equation}
\begin{equation}
\label{eq: energy splitting constraint}
    \beta_n^{\text{t}}, \beta_n^{\text{r}} \in [0, 1], \beta_n^{\text{t}} + \beta_n^{\text{r}}=1, 1\leq n\leq N,
\end{equation}
\begin{equation}
\label{eq: active beamforming constraint at BS}
    \sum_{k_{\text{l}}\in\{l_{\text{t}}, l_{\text{r}}\}}||\mathbf{w}_{k_{\text{l}}}||^2 \leq P_{\text{max}},
\end{equation}
\begin{equation}
\label{eq: MEs moving region}
  \mathbf{r}_n \in \mathcal{C}, 1 \leq n \leq N,
\end{equation}
\begin{equation}
\label{eq: distance between two MEs}
    ||\mathbf{r}_n-\mathbf{r}_{n'}||_2\geq d_0, 1 \leq n'\neq n \leq N,
\end{equation} 
\begin{equation}
\label{eq: legitimate user rate constraint}
    R_{k_{\text{l}}} \geq R^{\text{l}}_{\min},
\end{equation}
\begin{equation}
\label{eq: eve rate constraint}
    R_{k_{\text{e}}, k_{\text{l}}}^{\text{e}} \leq R^{\text{e}}_{\max}.
\end{equation}
\end{subequations}
Constraint~\eqref{eq: phase constraints} and~\eqref{eq: energy splitting constraint} provide feasible conditions of the STARS transmission and reflection coefficients. Constraint~\eqref{eq: MEs moving region} restricts the MEs movable region within $\mathcal{C}$. Constraint~\eqref{eq: distance between two MEs} guarantees that adjacent MEs distances do not exceed $D_0$ to avoid the coupling effect. Constraint \eqref{eq: active beamforming constraint at BS} ensures that the BS transmit power does not exceed the maximum value $P_{\text{max}}$. 
Constraint \eqref{eq: legitimate user rate constraint} ensures that the data rate for legitimate users is no less than the minimum data rate requirement $R^{\text{l}}_{\min}$. Constraint~\eqref{eq: eve rate constraint} ensures that the data rate for eavesdroppers does not exceed the maximum wiretapping rate $R^{\text{e}}_{\max}$. The optimization problem~\eqref{eq: optimization problem} is challenging to solve due to its non-convexity with respect to (w.r.t.) $\mathbf{R}$, $\mathbf{W}$, $\boldsymbol{\Theta}_{\text{t}}$ and $\boldsymbol{\Theta}_{\text{r}}$, as well as the tight coupling among the involved variables. In general, there is no typical method for solving such optimization problem efficiently. In the following, we propose an iterative algorithm by applying the AO technique.

\begin{figure*}[ht]
    \begin{equation}
    \label{eq: F matrix}
           \mathbf{F}_{k}(\mathbf{r}_n) = \begin{pmatrix}
        e^{j\frac{2\pi}{\lambda}\left(\rho_{\text{S}, k}^1(\mathbf{r}_n) - \rho_{\text{S}, \text{in}}^1(\mathbf{r}_n)\right)} & \cdots & e^{j\frac{2\pi}{\lambda}\left(\rho_{\text{S}, k}^1(\mathbf{r}_n) - \rho_{\text{S}, \text{in}}^{L_{\text{BS}}}(\mathbf{r}_n)\right)} \\
        \vdots & \ddots & \vdots \\
        e^{j\frac{2\pi}{\lambda}\left(\rho_{\text{S}, p}^{L_{\text{S},k}}(\mathbf{r}_n) - \rho_{\text{S}, \text{in}}^1(\mathbf{r}_n)\right)} & \cdots & e^{j\frac{2\pi}{\lambda}\left(\rho_{\text{S}, k}^{L_{\text{S},k}}(\mathbf{r}_n) - \rho_{\text{S}, \text{in}}^{L_{\text{BS}}}(\mathbf{r}_n)\right)}
        \end{pmatrix}
        , \forall k\in\mathcal{K}.
    \end{equation}
    \hrulefill
\end{figure*}

\section{Alternative Optimization Algorithm}
In this section, we propose an AO-based iterative algorithm for solving the intractable problem~\eqref{eq: optimization problem}. 
Specifically, for given active beamforming $\mathbf{W}$ and passive beamforming $\boldsymbol{\Theta}_{\varrho}$, we optimize the MEs positions $\mathbf{R}$ by invoking the penalty-based gradient ascent algorithm.
For any given MEs positions $\mathbf{R}$ and passive beamforming $\boldsymbol{\Theta}_{\varrho}$ (active beamforming $\mathbf{W}$), we proceed to optimize $\mathbf{W}$ ($\boldsymbol{\Theta}_{\varrho}$) using the SCA technique.
Then, we present the overall algorithm and analyze the convergence and complexity properties. 

\subsection{ME Position Optimization}

In this subproblem, we aim to optimize $\mathbf{R}$ while keeping $\boldsymbol{\Theta}_{\varrho}$, and $\mathbf{W}$ fixed. To begin with, we first reformulate problem~\eqref{eq: optimization problem} into a more tractable form. By introducing the STARS transmission and reflection coefficient vectors as $\mathbf{q}_{\varrho} = \left[\sqrt{\beta_1^{\varrho}}e^{j\theta_1^{\varrho}},\sqrt{\beta_2^{\varrho}}e^{j\theta_2^{\varrho}},\ldots, \sqrt{\beta_N^{\varrho}}e^{j\theta_N^{\varrho}}\right]^H$, $\forall \varrho\in\{\text{t}, \text{r}\}$, $\left|\mathbf{h}_{k_{\text{l}}}(\mathbf{R})\mathbf{\Theta}_{\varrho}\mathbf{H}(\mathbf{R})\mathbf{w}_{k_{\text{l}}}\right|^2$ and $\left|\mathbf{g}_{k_{\text{e}}}(\mathbf{R})\mathbf{\Theta}_{\varrho}\mathbf{H}(\mathbf{R})\mathbf{w}_{k_{\text{l}}}\right|^2$ can be rewritten as $\left|\mathbf{q}_{\varrho}^H\mathbf{V}_{k_{\text{l}}}(\mathbf{R})\mathbf{w}_{k_{\text{l}}}\right|^2$ and $\left|\mathbf{q}_{\varrho}^H\mathbf{V}_{k_{\text{e}}}(\mathbf{R})\mathbf{w}_{k_{\text{l}}}\right|^2$, respectively, where $\mathbf{V}_{k_{\text{l}}}(\mathbf{R})= \text{diag}\left(\mathbf{h}_{k_{\text{l}}}\left(\mathbf{R}\right)\right)\mathbf{H}(\mathbf{R}) \in \mathbb{C}^{N \times M}$ and $\mathbf{V}_{k_{\text{e}}}(\mathbf{R})= \text{diag}\left(\mathbf{g}_{k_{\text{e}}}(\mathbf{R})\right)\mathbf{H}(\mathbf{R}) \in \mathbb{C}^{N \times M}$. Additionally, we define $\mathbf{F}_k(\mathbf{r}_n) = \mathbf{f}_k(\mathbf{r}_n)\left(\mathbf{f}_{\text{in}}(\mathbf{r}_n)\right)^H \in \mathbb{C}^{L_{\text{S},k} \times L_{\text{BS}}}$, $\forall k \in \mathcal{K}$, which depend solely on the position of the $n$-th ME and can be expanded as in \eqref{eq: F matrix} at the top of this page.
Consequently, the expressions for $\left|\mathbf{q}_{\varrho}^H\mathbf{V}_k(\mathbf{R})\mathbf{w}_{k_{\text{l}}}\right|^2$ can be written as

\begin{equation}
\label{eq: element-wise of effective power for legitimate user}
    \left|\mathbf{q}_{\varrho}^H\mathbf{V}_k(\mathbf{R})\mathbf{w}_{k_{\text{l}}}\right|^2 = \left|\sum_{n=1}^N \mathbf{x}_{n,k}\mathbf{F}_k(\mathbf{r}_n)\mathbf{y}_{k_{\text{l}}}\right|^2, \forall k\in\mathcal{K},
\end{equation}
where $\mathbf{x}_{n,k} = \mathbf{q}_{\varrho}^H[n]\mathbf{\sigma}_{\mathbf{S}, k} \in \mathbb{C}^{1 \times L_{\text{S},k}}$, and $\mathbf{y}_{k_{\text{l}}} = \mathbf{\Sigma}_{\text{BS}}\mathbf{E}\mathbf{w}_{k_{\text{l}}} \in \mathbb{C}^{L_{\text{BS}} \times 1}$. Then, the secrecy rate of the legitimate user $k_{\text{l}}$ can be expressed as 

\begin{equation}
    \begin{split}
    \label{eq: secrecy rate function}
    R^{\text{s}}_{k_{\text{l}}} &= \left[\text{log}_2\left(1+\frac{\left|\sum_{n=1}^N \mathbf{x}_{n,k_{\text{l}}}\mathbf{F}_{k_{\text{l}}}(\mathbf{r}_n)\mathbf{y}_{k_{\text{l}}}\right|^2}{\left|\sum_{n=1}^N \mathbf{x}_{n,k_{\text{l}}}\mathbf{F}_{k_{\text{l}}}(\mathbf{r}_n)\mathbf{y}_{\bar{k}_{\text{l}}}\right|^2 + \sigma_{\text{l}}^2}\right) \right.\\
    -& \left. \max_{k_{\text{e}} \in \mathcal{K}_{\text{e}}}\text{log}_2\left(1+\frac{\left|\sum_{n=1}^N \mathbf{x}_{n,k_{\text{e}}}\mathbf{F}_{k_{\text{e}}}(\mathbf{r}_n)\mathbf{y}_{k_{\text{l}}}\right|^2}{\left|\sum_{n=1}^N \mathbf{x}_{n,k_{\text{e}}}\mathbf{F}_{k_{\text{e}}}(\mathbf{r}_n)\mathbf{y}_{\bar{k}_{\text{l}}}\right|^2 + \sigma_{\text{e}}^2}\right) \right]^+.
    \end{split}
\end{equation}
Based on~\eqref{eq: secrecy rate function}, the objective function can be expressed as a function solely dependent on $\mathbf{R}$. Thus, the original optimization problem~\eqref{eq: optimization problem} can be rewritten as   
\begin{subequations}
\label{eq:optimization problem r-1}
\begin{equation}
\label{eq: secrecy rate r-1}
    \max_{\mathbf{R}} g\left(\mathbf{R}\right) \triangleq \sum_{k_{\text{l}}\in\mathcal{K}_{\text{l}}} R^{\text{s}}_{k_{\text{l}}} 
\end{equation}
\begin{equation}
\label{eq: optimization problem r-1 constraint}
    {\rm{s.t.}} \ \  \eqref{eq: MEs moving region} \sim \eqref{eq: eve rate constraint}. \notag
\end{equation}
\end{subequations}
Problem~\eqref{eq:optimization problem r-1} is highly non-convex w.r.t. $\mathbf{R}$, making it difficult to obtain the global optimal solution. To address this, we propose to utilize the gradient ascent framework for the optimization of $\mathbf{R}$. 

Since the gradient ascent algorithm cannot directly solve the constrained optimization problem, we first transform~\eqref{eq:optimization problem r-1} into an unconstrained one. Specifically, to satisfy the movable region constraint of the MEs, as given in constraint~\eqref{eq: MEs moving region}, we introduce an auxiliary variable $\dot{\mathbf{R}} \in \mathbb{R}^{2 \times N}$ which satisfies
\begin{equation}
\label{eq: replacing the moving region}
    \mathbf{R} = \frac{A}{2}\tanh{\left(\dot{\mathbf{R}}\right)},
\end{equation}
where $\tanh{(x)} = \left(e^x - e^{-x}\right)/\left(e^x + e^{-x}\right) \in (-1,1)$, and $\dot{\mathbf{R}} = \left[\dot{\mathbf{r}}_1, \dot{\mathbf{r}}_2, \cdots, \dot{\mathbf{r}}_N \right]$. This operation projects the variables $\dot{\mathbf{r}}_n$, $\forall n$, which are initially defined in the real space, onto the confined real space $\mathcal{C}$. By substituting the relationship form~\eqref{eq: replacing the moving region} into problem~\eqref{eq:optimization problem r-1}, the problem~\eqref{eq:optimization problem r-1} can be rewritten as 
\begin{subequations}
\label{eq:optimization problem r-2}
    \begin{equation}
\label{eq: objective function rewritten r-1}
    \max_{\dot{\mathbf{R}}} g\left(\dot{\mathbf{R}}\right)
\end{equation}
\begin{equation}
\label{eq: distance between two MEs by replacing the variable}
\begin{split}
    {\rm{s.t.}} \ \  f_{n,n'}\left(\dot{\mathbf{R}}\right) ={d_0} - \frac{A}{2}&||\tanh{(\dot{\mathbf{r}}_n)}-\tanh{(\dot{\mathbf{r}}_{n'})}||_2 \leq 0, \\
    & \forall 1 \leq n \neq n' \leq N,
\end{split}
\end{equation} 
\begin{equation}
\label{eq: legitimate user rate constraint by replacing the variable}
    f_{k_{\text{l}}}\left(\dot{\mathbf{R}}\right) = R^{\text{l}}_{\min} - R_{k_{\text{l}}} \leq 0, \forall k_{\text{l}}\in\mathcal{K}_{\text{l}}, 
\end{equation}

\begin{equation}
\label{eq: eve rate constraint by replacing the variable}
    f_{k_{\text{e}}, k_{\text{l}}}\left(\dot{\mathbf{R}}\right) = -\left(R^{\text{e}}_{\max} - R_{k_{\text{e}}, k_{\text{l}}}^{\text{e}}\right) \leq 0, \forall k_{\text{e}}\in\mathcal{K}_{\text{e}}. 
\end{equation}
\end{subequations} 
We define $\mathcal{D}_N = \left\{(n, n')|n, n'\in\mathcal{N}, n\neq n'\right\}$ as the set of the minimum distance constraints between MEs. Additionally, we denote $\mathcal{I} = \mathcal{D}_N \cup \mathcal{K}_{\text{l}} \cup \mathcal{K}_{\text{e}}$ as the set of all inequality constraints. With this notation, the optimization  problem~\eqref{eq:optimization problem r-2} can be expressed as
\begin{subequations}
\label{eq: optimization problem r-3}
\begin{equation}
\label{eq: secrecy rate r-3}
    \max_{\dot{\mathbf{R}}} g\left(\dot{\mathbf{R}}\right) 
\end{equation}
\begin{equation}
\label{eq: inequality constraint}
    {\rm{s.t.}} \ \  f_i\left(\dot{\mathbf{R}}\right) \leq 0, \forall i \in \mathcal{I}.
\end{equation}
\end{subequations}

To handle the inequality constraint~\eqref{eq: inequality constraint}, a common approach is to use the penalty method~\cite{nocedal2006numerical}. This method involves adding a weighted penalty term, $\rho\sum_{i\in\mathcal{I}}\max\left\{0, f_i\left(\dot{\mathbf{R}}\right)\right\}$, to the objective function. Here, $\rho > 0$ is a penalty factor that determines the weight of the penalty applied when the constraint~\eqref{eq: inequality constraint} is violated. While effective in many cases, the considered penalty term becomes intractable due to its non-differentiability. To overcome this issue, we advocate to replace the max function with the continuous and differentiable log-sum-exp (LSE) function by exploiting the following approximation:
\begin{equation}
\label{eq: LSE expression}
    \max\left\{0, f_i\left(\dot{\mathbf{R}}\right)\right\} \approx \mu\log\left(1 + e^{f_i\left(\dot{\mathbf{R}}\right)/\mu}\right), 
\end{equation}
where $\mu > 0$ represents the smoothing parameter. The approximation in~\eqref{eq: LSE expression} becomes more precise as $\mu$ decreases. Based on~\eqref{eq: LSE expression}, the optimization problem~\eqref{eq: optimization problem r-3} can be equivalently transformed into the following unconstrained form 
\begin{equation}
\label{eq: secrecy rate r-4}
    \max_{\dot{\mathbf{R}}} G\left(\dot{\mathbf{R}}\right) = g\left(\dot{\mathbf{R}}\right) - \rho\sum_{i\in\mathcal{I}}\mu\log\left(1 + e^{f_i\left(\dot{\mathbf{R}}\right)/\mu}\right).
\end{equation}

We employ the gradient ascent algorithm for solving the problem~\eqref{eq: secrecy rate r-4}. Specifically, let $\alpha^{(\tau_1)}$ denote the learning rate, which determines the step size for updating the parameters during the $\tau_1$-th iteration. In the $\tau_1$-th iteration, the gradient $\nabla G\left(\dot{\mathbf{R}}^{(\tau_1)}\right) \in \mathbb{R}^{2 \times N}$ of the objective function $G\left(\dot{\mathbf{R}}\right)$ is computed at the current point $\dot{\mathbf{R}}^{(\tau_1)}$. Then, the parameter $\dot{\mathbf{R}}^{\left(\tau_1 + 1\right)}$ can be updated according to the following update rule: 
\begin{equation}
\label{eq: update rule}
    \dot{\mathbf{R}}^{(\tau_1 + 1)} = \dot{\mathbf{R}}^{(\tau_1)} + \alpha^{(\tau_1)}\nabla G\left(\dot{\mathbf{R}}^{(\tau_1)}\right).
\end{equation}
In~\eqref{eq: update rule}, the gradient vector $\nabla G\left(\dot{\mathbf{R}}^{(\tau_1)}\right)$ is calculated as 
\begin{equation}
\label{eq: gradient vector}
    \nabla G\left(\dot{\mathbf{R}}^{(\tau_1)}\right) = \nabla G\left(\mathbf{R}^{(\tau_1)}\right) \circ \frac{A}{2}\left(1 - \tanh^2{\left(\dot{\mathbf{R}}^{(\tau_1)}\right)}\right),
\end{equation}
where $\circ$ denotes the Hadamard multiplication. 
The detailed derivative of $\nabla G\left(\mathbf{R}\right)$ is provided in Appendix~\ref{app: A}. To determine the appropriate value of the learning rate $\alpha^{(\tau_1)}$, we employ the backtracking line search method~\cite{boyd2004convex}. This process begins with a relatively large initial value for $\alpha^{(\tau_1)}$  which is iteratively reduced by a factor of $\omega_\alpha$, such that $\alpha^{(\tau_1)} = \omega_\alpha \alpha^{(\tau_1)}$, until the Armijo condition is satisfied:
\begin{equation}
\label{eq: Armijo condition}
    G\left(\dot{\mathbf{R}}^{(\tau_1 + 1)}\right) \geq G\left(\dot{\mathbf{R}}^{(\tau_1)}\right) + \delta\alpha^{(\tau_1)} \left\Vert\nabla G\left(\dot{\mathbf{R}}^{(\tau_1)}\right)\right\Vert^2,
\end{equation}
where $\delta \in (0, 1)$ is a predefined parameter that governs the acceptable range of improvement in the objective function. The Armijo condition ensures that the step size is neither too large (risking overshooting) nor too small (slowing down the convergence speed). 

Note that, the efficacy of solving~\eqref{eq: secrecy rate r-4} to achieve an optimal solution for~\eqref{eq: optimization problem r-3} is intrinsically relevant to the selection of the penalty factor $\rho$ and the smoothing parameter $\mu$. Specifically, when $\rho$ is sufficiently large and $\mu$ is sufficiently small, problems~\eqref{eq: optimization problem r-3} and~\eqref{eq: secrecy rate r-4} become equivalent. {However, to prevent the solution from being overly influenced by the penalty term, $\rho$ should not be set too large initially.
Building upon this, the algorithm employs a two-layer iterative framework, as outlined in~\textbf{Algorithm~\ref{alg1: MEs positioning}}.  Specifically, in the outer layer, $\rho$ is updated by a factor $\omega_\rho>1$ with $\rho =\omega_\rho\rho$, while $\mu$ is updated by a factor $0 <\omega_\mu< 1$ with $\mu =\omega_\mu\mu$, until the constraints specified in equation~\eqref{eq: inequality constraint} are satisfied.} In the inner loop, the gradient ascent algorithm is applied on the problem~\eqref{eq: secrecy rate r-4} with the given $\rho$ and $\mu$, until the increment of $G\left(\dot{\mathbf{R}}\right)$ is below the predefined threshold $\varepsilon_1$ or the maximum number of iterations $\tau_1^{\max}$ is reached. 

\begin{algorithm}
	\renewcommand{\algorithmicrequire}{\textbf{Input:}}
	\renewcommand{\algorithmicensure}{\textbf{Output:}}
	\caption{Proposed gradient ascent algorithm for solving MEs positions subproblem}
	\label{alg1: MEs positioning}
	\begin{algorithmic}[1]
		\State Initialize points $\dot{\mathbf{R}}^{(0)} = \text{artanh}\left(\frac{2}{A}\mathbf{R}^{(0)}\right)$;
        \State Initialize penalty factor $\rho$;
        \State Initialize smoothing parameter $\mu$;
        \State Obtain $G\left(\dot{\mathbf{R}}\right)$;
		\Repeat
            \State Set $\tau_1=0$;
            \Repeat
                \State Calculate gradient $\nabla G\left(\dot{\mathbf{R}}^{(\tau_1)}\right)$;
                \State Initialize step size $\alpha^{(\tau_1)} = \alpha^{(0)}$;
                \While {Armijo condition \eqref{eq: Armijo condition} is not satisfied} \label{line:armijo}
                    \State Shrink the step size $\alpha^{(\tau_1)} = \omega_\alpha \alpha^{(\tau_1)}$;
                    \State Update MEs position vector in \eqref{eq: update rule};
                \EndWhile
                \State $\tau_1 = \tau_1 + 1$;
            \Until $\left|G\left(\dot{\mathbf{R}}^{(\tau_1+1)}\right) - G\left(\dot{\mathbf{R}}^{(\tau_1)}\right)\right| \leq \varepsilon_1$ \textbf{or} $\tau_1 \geq \tau_1^{\max}$;
            \State Update $\rho = \omega_\rho \rho$, $\mu = \omega_\mu \mu$;
        \Until $\frac{A}{2} || \tanh{\left(\dot{\mathbf{r}}_n\right)} - \tanh{\left(\dot{\mathbf{r}}_{n'}\right)} ||_2 \geq d_0$, $R_{k_{\text{l}}} \geq R^{\text{l}}_{\min}$, $R_{k_{\text{e}}, k_{\text{l}}}^{\text{e}} \leq R^{\text{e}}_{\max}$.
        
        \State \Return $\mathbf{R} = \frac{A}{2} \tanh{\left(\dot{\mathbf{R}}\right)}$.
	\end{algorithmic}  
\end{algorithm}

\subsection{STARS Passive Beamforming Optimization}
With given $\mathbf{R}$, and $\mathbf{W}$, we first need to address the challenges proposed by the quadratic forms for the optimization of $\boldsymbol{\Theta}_{\varrho}$. Specifically, we set $\mathbf{H}_{k_{\text{l}}, k_{\text{l}}} = \mathbf{V}_{k_{\text{l}}}(\mathbf{R})\mathbf{w}_{k_{\text{l}}} \left(\mathbf{V}_{k_{\text{l}}}(\mathbf{R})\mathbf{w}_{k_{\text{l}}}\right)^H$, $\forall k_{\text{l}} \in \mathcal{K}_{\text{l}}$, and $\mathbf{H}_{k_{\text{e}}, k_{\text{l}}} = \mathbf{V}_{k_{\text{e}}}(\mathbf{R})\mathbf{w}_{k_{\text{l}}} \left(\mathbf{V}_{k_{\text{e}}}(\mathbf{R})\mathbf{w}_{k_{\text{l}}}\right)^H, \forall k_{\text{e}} \in \mathcal{K}_{\text{e}}$. Moreover, define $\mathbf{\Phi}_{\varrho} = \mathbf{q}_{\varrho}(\mathbf{q}_{\varrho})^H, \forall \varrho \in \{\text{t}, \text{r}\}$, which satisfy $\text{rank}\left(\mathbf{\Phi}_{\varrho}\right)=1$, $\mathbf{\Phi}_\varrho \succeq 0$, and $\text{diag}\left(\mathbf{\Phi}_{\varrho}\right) = \boldsymbol{\beta}_\varrho = \left[\beta_1^\varrho, \beta_2^\varrho, \ldots, \beta_N^\varrho\right]$. Thus, $R^{\text{s}}_{k_{\text{l}}}$ can be rewritten as
\begin{equation}
    \begin{split}
    \label{eq: objective function rewritten by W}
    R^{\text{s}}_{k_{\text{l}}} = &\left[\text{log}_2\left(1+\frac{ \text{Tr}\left(\mathbf{H}_{k_{\text{l}}, k_{\text{l}}}\mathbf{\Phi}_{\varrho}\right)}{\text{Tr}\left(\mathbf{H}_{k_{\text{l}}, \bar{k}_{\text{l}}}\mathbf{\Phi}_{\varrho}\right) + \sigma_{\text{l}}^2}\right) \right.\\
    &- \left. \max_{k_{\text{e}} \in \mathcal{K}_{\text{e}}}\text{log}_2\left(1+\frac{\text{Tr}\left(\mathbf{H}_{k_{\text{e}}, k_{\text{l}}}\mathbf{\Phi}_\varrho \right)}{\text{Tr}\left(\mathbf{H}_{k_{\text{e}}, \bar{k}_{\text{l}}}\mathbf{\Phi}_{\varrho}\right) + \sigma_{\text{e}}^2}\right) \right]^+.
    \end{split}
\end{equation}

Subsequently, with the given positions of the MEs and the active beamforming of the BS, the subproblem of optimizing the STARS passive beamforming can be formulated as
\begin{subequations}
\label{eq: optimization problem C-1}
\begin{equation}
\label{eq: secrecy rate C-1}
    \max_{\boldsymbol{\Phi}_\varrho} \sum_{k_{\text{l}}\in\mathcal{K}_{\text{l}}} R^{\text{s}}_{k_{\text{l}}} 
\end{equation}
\begin{equation}
\label{eq: diagonal constraint}
    {\rm{s.t.}} \ \ \text{diag}(\mathbf{\Phi}_{\varrho}) = \boldsymbol{\beta}_\varrho,
\end{equation}
\begin{equation}
\label{eq: positive semidefinite matrix}
    \mathbf{\Phi}_\varrho \succeq 0,
\end{equation}
\begin{equation}
\label{eq: rank one constraint}
    \text{rank}(\mathbf{\Phi}_{\varrho})=1,
\end{equation}
\begin{equation}
\label{eq: optimization before constraint A}
    \eqref{eq: phase constraints}, \eqref{eq: energy splitting constraint}, \eqref{eq: legitimate user rate constraint}, \eqref{eq: eve rate constraint}, \notag
\end{equation}
\end{subequations}
where $\varrho\in\{\text{t}, \text{r}\}$. The objective function~\eqref{eq: objective function rewritten by W} and constraints \eqref{eq: rank one constraint}, \eqref{eq: legitimate user rate constraint} and \eqref{eq: eve rate constraint} are non-convex w.r.t. $\boldsymbol{\Phi}_\varrho$. To address the non-convexity of \eqref{eq: objective function rewritten by W}, we introduce auxiliary variables $\left\{A^{\text{S}}_{k_{\text{l}}}, B^{\text{S}}_{k_{\text{l}}}, C^{\text{S}}_{k_{\text{l}}}, D^{\text{S}}_{k_{\text{e}}, k_{\text{l}}}, E^{\text{S}}_{k_{\text{e}}, k_{\text{l}}}, F^{\text{S}}_{k_{\text{e}}, k_{\text{l}}} \right\}$ to construct a liner expression of \eqref{eq: objective function rewritten by W}. Specifically, consider the following variables substitution:
\begin{equation}
\label{eq: legitimate rate numerator}
    2^{A^{\text{S}}_{k_{\text{l}}}} = \text{Tr}\left(\mathbf{H}_{k_{\text{l}}, k_{\text{l}}}\mathbf{\Phi}_{\varrho} \right) +\text{Tr}\left(\mathbf{H}_{k_{\text{l}}, \bar{k}_{\text{l}}}\mathbf{\Phi}_{\varrho} \right) + \sigma_{\text{l}}^2,
\end{equation}

\begin{equation}
\label{eq: legitimate rate denominator}
    C^{\text{S}}_{k_{\text{l}}} = 2^{B^{\text{S}}_{k_{\text{l}}}} = \text{Tr}\left(\mathbf{H}_{k_{\text{l}}, \bar{k}_{\text{l}}}\mathbf{\Phi}_{\varrho} \right) + \sigma_{\text{l}}^2.
\end{equation}
Then, we have $\widehat{R}_{k_{\text{l}}} = A^{\text{S}}_{k_{\text{l}}} - B^{\text{S}}_{k_{\text{l}}}$. Similarly, we define $\bar{R}^{\text{e}}_{k_{\text{e}}, k_{\text{l}}} = \left\{ D^{\text{S}}_{k_{\text{e}}, k_{\text{l}}} - E^{\text{S}}_{k_{\text{e}}, k_{\text{l}}} \right\}$, where the auxiliary variables $D^{\text{S}}_{k_{\text{e}}, k_{\text{l}}}$, $E^{\text{S}}_{k_{\text{e}}, k_{\text{l}}}$ and $F^{\text{S}}_{k_{\text{e}}, k_{\text{l}}}$ satisfy
\begin{equation}
\label{eq: eve rate denominator}
   F^{\text{S}}_{k_{\text{e}}, k_{\text{l}}} = 2^{D^{\text{S}}_{k_{\text{e}}, k_{\text{l}}}} = \text{Tr}\left(\mathbf{H}_{k_{\text{e}}, k_{\text{l}}}\mathbf{\Phi}_{\varrho} \right) + \text{Tr}\left(\mathbf{H}_{k_{\text{e}}, \bar{k}_{\text{l}}}\mathbf{\Phi}_{\varrho} \right) + \sigma_{\text{e}}^2,
\end{equation}

\begin{equation}
\label{eq: eve rate numerator}
    2^{E^{\text{S}}_{k_{\text{e}}, k_{\text{l}}}} = \text{Tr}\left(\mathbf{H}_{k_{\text{e}}, \bar{k}_{\text{l}}}\mathbf{\Phi}_{\varrho} \right) + \sigma_{\text{e}}^2.
\end{equation}

Based on~\eqref{eq: legitimate rate numerator} - \eqref{eq: eve rate numerator}, ${R}^s_{k_{\text{l}}}$ can be {rewritten as}
\begin{equation}
\label{eq: secrecy rate replace}
    {R}^s_{k_{\text{l}}} = \bar{R}^s_{k_{\text{l}}} = \left[\widehat{R}_{k_{\text{l}}} - \max\left\{\bar{R}^{\text{e}}_{e_{\text{t}}, k_{\text{l}}}, \bar{R}^{\text{e}}_{e_{\text{r}}, k_{\text{l}}}\right\}\right]^+, \forall k_{\text{l}}\in\mathcal{K}_{\text{l}}.
\end{equation}
By introducing auxiliary variables and applying linearization, the original non-convex objective function~\eqref{eq: objective function rewritten by W} is transformed into the convex form~\eqref{eq: secrecy rate replace}. At the same time, inequality constraints are imposed on the optimization variables $\left\{A^{\text{S}}_{k_{\text{l}}}, B^{\text{S}}_{k_{\text{l}}}, C^{\text{S}}_{k_{\text{l}}}, D^{\text{S}}_{k_{\text{e}}, k_{\text{l}}}, E^{\text{S}}_{k_{\text{e}}, k_{\text{l}}}, F^{\text{S}}_{k_{\text{e}}, k_{\text{l}}}\right\}$. As a result, the optimization problem can be reformulated as

\begin{subequations}
\label{eq: optimization problem C-2}
\begin{equation}
\label{eq: secrecy rate C-3}
    \max_{\left\{\boldsymbol{\Phi}_\varrho, A^{\text{S}}_{k_{\text{l}}}, B^{\text{S}}_{k_{\text{l}}}, C^{\text{S}}_{k_{\text{l}}}, D^{\text{S}}_{k_{\text{e}}, k_{\text{l}}}, E^{\text{S}}_{k_{\text{e}}, k_{\text{l}}}, F^{\text{S}}_{k_{\text{e}}, k_{\text{l}}} \right\}} \sum_{k_{\text{l}}\in\mathcal{K}_{\text{l}}} \bar{R}^s_{k_{\text{l}}}  
\end{equation}
\begin{equation}
\label{eq: legitimate rate numerator constraint}
    {\rm{s.t.}} \ \  2^{A^{\text{S}}_{k_{\text{l}}}} \leq \text{Tr}\left(\mathbf{H}_{k_{\text{l}}, k_{\text{l}}}\mathbf{\Phi}_{\varrho} \right) +\text{Tr}\left(\mathbf{H}_{k_{\text{l}}, \bar{k}_{\text{l}}}\mathbf{\Phi}_{\varrho} \right) + \sigma_{\text{l}}^2,
\end{equation}
\begin{equation}
\label{eq: legitimate rate middle constraint}
   C^{\text{S}}_{k_{\text{l}}} \geq \text{Tr}\left(\mathbf{H}_{k_{\text{l}}, \bar{k}_{\text{l}}}\mathbf{\Phi}_{\varrho} \right) + \sigma_{\text{l}}^2,
\end{equation}
\begin{equation}
\label{eq: eve rate numerator constraint}
   F^{\text{S}}_{k_{\text{e}}, k_{\text{l}}} \geq \text{Tr}\left(\mathbf{H}_{k_{\text{e}}, k_{\text{l}}}\mathbf{\Phi}_{\varrho} \right) + \text{Tr}\left(\mathbf{H}_{k_{\text{e}}, \bar{k}_{\text{l}}}\mathbf{\Phi}_{\varrho} \right) + \sigma_{\text{e}}^2.
\end{equation}
\begin{equation}
\label{eq: eve rate middle constraint}
    2^{E^{\text{S}}_{k_{\text{e}}, k_{\text{l}}}} \leq \text{Tr}\left(\mathbf{H}_{k_{\text{e}}, \bar{k}_{\text{l}}}\mathbf{\Phi}_{\varrho} \right) + \sigma_{\text{e}}^2,
\end{equation}
\begin{equation}
\label{eq: legitimate rate denominator constraint}
   2^{B^{\text{S}}_{k_{\text{l}}}} \geq C^{\text{S}}_{k_{\text{l}}},
\end{equation}
\begin{equation}
\label{eq: eve rate denominator constraint}
    2^{D^{\text{S}}_{k_{\text{e}}, k_{\text{l}}}} \geq F^{\text{S}}_{k_{\text{e}}, k_{\text{l}}},
\end{equation}
\begin{equation}
\label{eq: legitimate user rate constraint C-3}
     \widehat{R}_{k_{\text{l}}} \geq R^{\text{l}}_{\min},
\end{equation}
\begin{equation}
\label{eq: eve rate constraint C-3}
    \bar{R}^{\text{e}}_{k_{\text{e}}, k_{\text{l}}} \leq R^{\text{e}}_{\max},
\end{equation}
\begin{equation}
\label{eq: optimization before constraint A}
    \\ \eqref{eq: phase constraints}, \eqref{eq: energy splitting constraint}, \eqref{eq: diagonal constraint}\sim\eqref{eq: rank one constraint}. \notag
\end{equation}
\end{subequations}

In problem~\eqref{eq: optimization problem C-2}, the constraints \eqref{eq: legitimate rate denominator constraint}, \eqref{eq: eve rate denominator constraint} and~\eqref{eq: rank one constraint} remain non-convex. For \eqref{eq: legitimate rate denominator constraint} and \eqref{eq: eve rate denominator constraint}, the first-order Taylor approximations in the $\tau_2$-iteration of the SCA are respectively given by
\begin{equation}
\label{eq: legitimate rate denominator constraint first-order Taylor}
   B^{\text{S}}_{k_{\text{l}}} \geq \text{log}_2\left(C^{\text{S}(\tau_2)}_{k_{\text{l}}}\right) + \frac{C^{\text{S}}_{k_{\text{l}}} - C^{\text{S}(\tau_2)}_{k_{\text{l}}}}{C^{\text{S}(\tau_2)}_{k_{\text{l}}} \text{ln}2},
\end{equation}

\begin{equation}
\label{eq: eve rate numerator constraint first-order Taylor}
   D^{\text{S}}_{k_{\text{e}}, k_{\text{l}}} \geq \text{log}_2\left( F^{\text{S}(\tau_2)}_{k_{\text{e}}, k_{\text{l}}} \right) + \frac{F^{\text{S}}_{k_{\text{e}}, k_{\text{l}}} - F^{\text{S}(\tau_2)}_{k_{\text{e}}, k_{\text{l}}}}{F^{\text{S}(\tau_2)}_{k_{\text{e}}, k_{\text{l}}} \text{ln}2},
\end{equation}
where $C^{\text{S}(\tau_2)}_{k_{\text{l}}}$ and $D^{\text{S}(\tau_2)}_{k_{\text{e}}, k_{\text{l}}}$ are the given local points of $C^{\text{S}}_{k_{\text{l}}}$ and $D^{\text{S}}_{k_{\text{e}}, k_{\text{l}}}$, respectively, in the $\tau_2$-th iteration of the SCA. 

The only remaining non-convexity in problem \eqref{eq: optimization problem C-2} is the rank-one constraint \eqref{eq: rank one constraint}, which can be tackled by invoking the penalty-based method~\cite{nocedal2006numerical}. Specifically, we first transform \eqref{eq: rank one constraint} into the following equivalent form
\begin{equation} 
\label{eq: equality constraint}
    \xi_\varrho^0 = \text{Tr}\left(\boldsymbol{\Phi}_{\varrho}\right) - ||\boldsymbol{\Phi}_{\varrho}||_2 = 0, \forall \varrho\in\{\text{t}, \text{r}\},
\end{equation}
where $||\boldsymbol{\Phi}_\varrho||_2 = \sigma_1(\boldsymbol{\Phi}_\varrho)$ represents the spectral norm, and $\sigma_1(\boldsymbol{\Phi}_\varrho)$ represents the largest singular value of the matrix $\boldsymbol{\Phi}_\varrho$. 
By incorporating \eqref{eq: equality constraint} as a penalty term into the objective function~\eqref{eq: secrecy rate C-3}, problem~\eqref{eq: optimization problem C-2} can be rewritten as
\begin{subequations}
\label{eq: optimization problem C-4}
\begin{equation}
\label{eq: secrecy rate C-4}
    \max_{\left\{\boldsymbol{\Phi}_\varrho, A^{\text{S}}_{k_{\text{l}}}, B^{\text{S}}_{k_{\text{l}}}, C^{\text{S}}_{k_{\text{l}}},  D^{\text{S}}_{k_{\text{e}}, k_{\text{l}}}, E^{\text{S}}_{k_{\text{e}}, k_{\text{l}}}, F^{\text{S}}_{k_{\text{e}}, k_{\text{l}}} \right\}} \sum_{k_{\text{l}}\in\mathcal{K}_{\text{l}}} \bar{R}^s_{k_{\text{l}}}  - \eta\sum_{\varrho\in\{\text{t},\text{r}\}} \xi_\varrho^0
\end{equation}
\begin{equation}
\label{eq: optimization before constraint C-4}
    {\rm{s.t.}} \ \  \eqref{eq: phase constraints}, \eqref{eq: energy splitting constraint}, \eqref{eq: diagonal constraint}, \eqref{eq: positive semidefinite matrix}, \eqref{eq: legitimate rate numerator constraint} \sim \eqref{eq: eve rate middle constraint}, \atop  \eqref{eq: legitimate user rate constraint C-3}, \eqref{eq: eve rate constraint C-3}, \eqref{eq: legitimate rate denominator constraint first-order Taylor}, \eqref{eq: eve rate numerator constraint first-order Taylor} \notag,
\end{equation}
\end{subequations}
where $\eta \geq 0$ represents the penalty factor. To handle the non-convexity of the penalty term, during the $\tau_2$-th iteration of the SCA, we obtain the convex upper bound by employing the first-order Taylor expansion at the point $\boldsymbol{\Phi}_\varrho^{(\tau_2)}$ as follows
\begin{equation} 
\label{eq: equality constraint 2}
    \text{Tr}(\boldsymbol{\Phi}_{\varrho}) - \left\Vert\boldsymbol{\Phi}_{\varrho}\right\Vert_2 \leq \text{Tr}(\boldsymbol{\Phi}_{\varrho}) -\bar{\boldsymbol{\Phi}}_\varrho^{(\tau_2)} \triangleq \xi_\varrho^{(\tau_2)},
\end{equation}
where $\bar{\boldsymbol{\Phi}}_\varrho^{(\tau_2)} \triangleq \left\Vert\boldsymbol{\Phi}_{\varrho}^{(\tau_2)}\right\Vert_2 + \text{Tr}\left[\bar{\mathbf{x}}\left(\boldsymbol{\Phi}_\varrho^{(\tau_2)}\right) \left(\bar{\mathbf{x}}\left(\boldsymbol{\Phi}_\varrho^{(\tau_2)}\right)\right)^H \left(\boldsymbol{\Phi}_\varrho - \boldsymbol{\Phi}_\varrho^{(\tau_2)}\right) \right]$, with $\bar{\mathbf{x}}\left(\boldsymbol{\Phi}_\varrho^{(\tau_2)}\right)$ represents the eigenvector associated with the largest eigenvalues of the matrix $\boldsymbol{\Phi}_\varrho^{(\tau_2)}$. Thus, the optimization problem \eqref{eq: optimization problem C-4} can be equivalently transformed into the following convex form
\begin{subequations}
\label{eq: optimization problem C-5}
\begin{equation}
\label{eq: secrecy rate}
    \max_{\left\{\boldsymbol{\Phi}_\varrho, A^{\text{S}}_{k_{\text{l}}}, B^{\text{S}}_{k_{\text{l}}}, C^{\text{S}}_{k_{\text{l}}}, D^{\text{S}}_{k_{\text{e}}, k_{\text{l}}}, E^{\text{S}}_{k_{\text{e}}, k_{\text{l}}}, F^{\text{S}}_{k_{\text{e}}, k_{\text{l}}} \right\}} \sum_{k_{\text{l}}\in\mathcal{K}_{\text{l}}} \bar{R}^s_{k_{\text{l}}}  - \eta\sum_{\varrho\in\{{\text{t}}, {\text{r}}\}} \xi_\varrho^{(\tau_2)}
\end{equation}
\begin{equation}
\label{eq: optimization before constraint C-5}
    {\rm{s.t.}} \ \  \eqref{eq: phase constraints}, \eqref{eq: energy splitting constraint}, \eqref{eq: diagonal constraint}, \eqref{eq: positive semidefinite matrix}, \eqref{eq: legitimate rate numerator constraint} \sim \eqref{eq: eve rate middle constraint}, \atop \eqref{eq: legitimate user rate constraint C-3}, \eqref{eq: eve rate constraint C-3}, \eqref{eq: legitimate rate denominator constraint first-order Taylor}, \eqref{eq: eve rate numerator constraint first-order Taylor}. \notag
\end{equation}
\end{subequations}
The relaxed problem \eqref{eq: optimization problem C-5} is a semi-definite program (SDP), which can be solved using the standard convex problem solvers, such as CVX~\cite{grant2014cvx}.
Note that, to avoid the penalty term dominating the objective function, $\eta$ should be initially set to a small value and gradually increased to ensure compliance with the rank-one condition. Therefore, we set a nested loop in the proposed algorithm for solving the original problem~\eqref{eq: optimization problem C-1}, as shown in~\textbf{Algorithm~\ref{alg1: STARS's Beamforming Vectors}}. In the inner loop, the variable $\{\boldsymbol{\Phi}_\varrho\}$ is iteratively updated by solving problem \eqref{eq: optimization problem C-5}, until the increment of the sum secrecy rate is below a threshold $\varepsilon_2$ or the number of inner iterations reaches the maximum number $\tau_2^{\max}$. In the outer loop, the penalty factor is gradually increased in each iteration according to the update rule $\eta = \omega_\eta\eta$, where $\omega_\eta > 1$. The outer loop terminates when the difference between $\text{Tr}(\boldsymbol{\Phi}_\varrho)$ and $||\boldsymbol{\Phi}_\varrho||_2$ becomes smaller than a given threshold $\epsilon_2$.
This iterative refinement ensures that the solution progressively satisfies the original problem's constraint~\eqref{eq: rank one constraint} as $\eta$ increases. By iteratively updating the penalty factor in the outer loop and optimizing the variables $\left\{\boldsymbol{\Phi}_\varrho, A^{\text{S}}_{k_{\text{l}}}, B^{\text{S}}_{k_{\text{l}}}, C^{\text{S}}_{k_{\text{l}}}, D^{\text{S}}_{k_{\text{e}}, k_{\text{l}}}, E^{\text{S}}_{k_{\text{e}}, k_{\text{l}}}, F^{\text{S}}_{k_{\text{e}}, k_{\text{l}}} \right\}$ in the inner loop, the penalty-based approach guarantees convergence to a feasible solution of the original problem as $\eta$ becomes sufficiently large. 
 
\begin{algorithm}
	\renewcommand{\algorithmicrequire}{\textbf{Input:}}
	\renewcommand{\algorithmicensure}{\textbf{Output:}}
	\caption{Proposed penalty-based SCA algorithm for solving STARS passive beamforming subproblem}
	\label{alg1: STARS's Beamforming Vectors}
	\begin{algorithmic}[1]
		\State Initialize feasible points $\left\{\boldsymbol{\Phi}_\varrho^{(0)}, A^{\text{S}(0)}_{k_{\text{l}}}, B^{\text{S}(0)}_{k_{\text{l}}}, C^{\text{S}(0)}_{k_{\text{l}}}, D^{\text{S}(0)}_{k_{\text{e}}, k_{\text{l}}}, E^{\text{S}(0)}_{k_{\text{e}}, k_{\text{l}}}, F^{\text{S}(0)}_{k_{\text{e}}, k_{\text{l}}} \right\}$;
        \State Initialize penalty factor $\eta$;
		\Repeat
            \State Set $\tau_2=0$;
                \Repeat
		      \State With the given points $\left\{\boldsymbol{\Phi}_\varrho^{(0)}, A^{\text{S}(0)}_{k_{\text{l}}}, B^{\text{S}(0)}_{k_{\text{l}}}, C^{\text{S}(0)}_{k_{\text{l}}}, D^{\text{S}(0)}_{k_{\text{e}}, k_{\text{l}}}, E^{\text{S}(0)}_{k_{\text{e}}, k_{\text{l}}}, F^{\text{S}(0)}_{k_{\text{e}}, k_{\text{l}}}\right\}$, solve the optimization problem \eqref{eq: optimization problem C-5};
		      \State Update $\left\{\boldsymbol{\Phi}_\varrho^{(\tau_2)}, A^{\text{S}(\tau_2)}_{k_{\text{l}}}, B^{\text{S}(\tau_2)}_{k_{\text{l}}}, C^{\text{S}(\tau_2)}_{k_{\text{l}}}, D^{\text{S}(\tau_2)}_{k_{\text{e}}, k_{\text{l}}} \right.$, $\left. E^{\text{S}(\tau_2)}_{k_{\text{e}}, k_{\text{l}}}, F^{\text{S}(\tau_2)}_{k_{\text{e}}, k_{\text{l}}} \right\}$ with current solution;
            \State $\tau_2=\tau_2+1$;
		\Until the fractional increase of the object value is below $\varepsilon_2$ \textbf{or} $\tau_2$ reaches the maximum number $\tau_2^{\text{max}}$;
            \State Update $\eta=\omega_\eta\eta$;
        \Until the penalty term value is below $\epsilon_2$.
        \State \Return $\boldsymbol{\Phi}_\varrho^*$.
	\end{algorithmic}  
\end{algorithm}

\subsection{BS Active Beamforming Optimization}
For the optimization of $\mathbf{W}$ with given 
$\mathbf{R}$, and $\boldsymbol{\Theta}_{\varrho}$, the subproblem can be formulated as
\begin{subequations}
\label{eq: optimization problem W-1}
\begin{equation}
\label{eq: optimal objective function W-1}
    \max_{\mathbf{W}} \sum_{k_{\text{l}} \in \mathcal{K}_{\text{l}}} R^{\text{s}}_{k_{\text{l}}} 
\end{equation}
\begin{equation}
\label{eq: W rank one constraint}
    {\rm{s.t.}} \ \  \text{Rank}(\mathbf{W}_{k_{\text{l}}})=1,
\end{equation}
\begin{equation}
\label{eq: W positive semidefinite matrix}
    \mathbf{W}_{k_{\text{l}}} \succeq 0,
\end{equation}
\begin{equation}
\label{eq: active beamforming constraint at BS W-1}
    \sum_{k_{\text{l}}\in\mathcal{K}_{\text{l}}}\text{Tr}(\mathbf{W}_{k_{\text{l}}}) \leq P_{\text{max}},
\end{equation}
\begin{equation}
\label{eq: optimization constraint W-1}
    \eqref{eq: legitimate user rate constraint}, \eqref{eq: eve rate constraint}, \notag
\end{equation}
\end{subequations}
where $\mathbf{W}_{k_{\text{l}}} = \mathbf{w}_{k_{\text{l}}} (\mathbf{w}_{k_{\text{l}}})^H$, $\forall k_{\text{l}} \in \mathcal{K}_{\text{l}}$. To address the non-convexity of \eqref{eq: optimal objective function W-1}, exponential slack variables $A_{k_{\text{l}}}$, $B_{k_{\text{l}}}$, $C_{k_{\text{e}}, k_{\text{l}}}$, and $D_{k_{\text{e}}, k_{\text{l}}}$ are introduced to derive a linear lower bound as follows:
\begin{equation}
\label{eq: legitimate rate numerator W}
    e^{A_{k_{\text{l}}}} \leq \text{Tr}\left(\mathbf{H}_{k_{\text{l}}}\mathbf{W}_{k_{\text{l}}} \right) +\text{Tr}\left(\mathbf{H}_{k_{\text{l}}}\mathbf{W}_{\bar{k}_{\text{l}}} \right) + \sigma_{\text{l}}^2,
\end{equation}

\begin{equation}
\label{eq: legitimate rate denominator W}
   e^{B_{k_{\text{l}}}} \geq \text{Tr}\left(\mathbf{H}_{k_{\text{l}}}\mathbf{W}_{\bar{k}_{\text{l}}} \right) + \sigma_{\text{l}}^2,
\end{equation}

\begin{equation}
\label{eq: eve rate denominator W}
   e^{C_{k_{\text{e}}, k_{\text{l}}}} \geq \text{Tr}\left(\mathbf{H}_{k_{\text{e}}}\mathbf{W}_{k_{\text{l}}} \right) + \text{Tr}\left(\mathbf{H}_{k_{\text{e}}}\mathbf{W}_{\bar{k}_{\text{l}}} \right) + \sigma_{\text{e}}^2,
\end{equation}

\begin{equation}
\label{eq: eve rate numerator W}
    e^{D_{k_{\text{e}}, k_{\text{l}}}} \leq \text{Tr}\left(\mathbf{H}_{k_{\text{e}}}\mathbf{W}_{\bar{k}_{\text{l}}} \right) + \sigma_{\text{e}}^2,
\end{equation}
where $\mathbf{H}_{k_{\text{l}}} = (\mathbf{q}_{\varrho}^H\mathbf{V}_{k_{\text{l}}})^H \mathbf{q}_{\varrho}^H\mathbf{V}_{k_{\text{l}}}$, and $\mathbf{H}_{k_{\text{e}}} = (\mathbf{q}_{\varrho}^H\mathbf{V}_{k_{\text{e}}})^H \mathbf{q}_{\varrho}^H\mathbf{V}_{k_{\text{e}}}$. Based on the inequalities \eqref{eq: legitimate rate numerator W} - \eqref{eq: eve rate numerator W}, the linear lower bound of ${R}^s_{k_{\text{l}}}$ can be expressed as 
\begin{equation}
\label{eq:linear lower bound}
    {R}^s_{k_{\text{l}}} \geq \tilde{R}^s_{k_{\text{l}}} = \left[\bar{R}_{k_{\text{l}}} - \max\left\{\tilde{R}^{\text{e}}_{e_{\text{t}}, k_{\text{l}}}, \tilde{R}^{\text{e}}_{e_{\text{r}}, k_{\text{l}}} \right\}\right]^+,
\end{equation}
where $\bar{R}_{k_{\text{l}}} = \left(A_{k_{\text{l}}} - B_{k_{\text{l}}}\right)/\ln2 \leq R_{k_{\text{l}}}$ and $\tilde{R}^{\text{e}}_{k_{\text{e}}, k_{\text{l}}} = \left(C_{k_{\text{e}}, k_{\text{l}}} - D_{k_{\text{e}}, k_{\text{l}}} \right)/\ln2 \geq R^{\text{e}}_{k_{\text{e}}, k_{\text{l}}}$. As a result, the optimization problem~\eqref{eq: optimization problem W-1} can be reformulated as  

\begin{subequations}
\label{eq: optimization problem W-2}
\begin{equation}
\label{eq: optimal objective function W-2}
    \max_{\left\{\mathbf{W}, A_{k_{\text{l}}}, B_{k_{\text{l}}}, C_{k_{\text{e}}, k_{\text{l}}}, D_{k_{\text{e}}, k_{\text{l}}} \right\}} \sum_{k_{\text{l}}\in\mathcal{K}_{\text{l}}} \tilde{R}^s_{k_{\text{l}}}
\end{equation}
\begin{equation}
\label{eq: legitimate user rate constraint W-2}
    {\rm{s.t.}} \ \  \bar{R}_{k_{\text{l}}} \geq R^{\text{l}}_{\min},
\end{equation}
\begin{equation}
\label{eq: eve rate constraint W-2}
    \tilde{R}^{\text{e}}_{k_{\text{e}}, k_{\text{l}}} \leq R^{\text{e}}_{\max},
\end{equation}
\begin{equation}
\label{eq: optimization before constraint W-2}
    \eqref{eq: W rank one constraint} \sim \eqref{eq: active beamforming constraint at BS W-1}, \eqref{eq: legitimate rate numerator W} \sim \eqref{eq: eve rate numerator W}. \notag
\end{equation}
\end{subequations}

The problem \eqref{eq: optimization problem W-2} remains non-convex due to the constraints \eqref{eq: legitimate rate denominator W}, \eqref{eq: eve rate denominator W}, and the rank-one constraint \eqref{eq: W rank one constraint}. For \eqref{eq: legitimate rate denominator W} and \eqref{eq: eve rate denominator W}, we construct the first-order Taylor expansions for their right-hand sides, which are given by
\begin{equation}
\label{eq: legitimate rate denominator constraint Taylor approximations}
  \text{Tr}\left(\mathbf{H}_{k_{\text{l}}}\mathbf{W}_{\bar{k}_{\text{l}}} \right) + \sigma_{\text{l}}^2 \leq e^{B_{k_{\text{l}}}^{(\tau_3)}} \left(B_{k_{\text{l}}} - B_{k_{\text{l}}}^{(\tau_3)} + 1\right),
\end{equation}

\begin{equation}
    \begin{split}
    \label{eq: eve rate denominator constraint Taylor approximations}
      \text{Tr}\left(\mathbf{H}_{k_{\text{e}}}\mathbf{W}_{k_{\text{l}}} \right) + &\text{Tr}\left(\mathbf{H}_{k_{\text{e}}}\mathbf{W}_{\bar{k}_{\text{l}}} \right) + \sigma_{\text{e}}^2 \\ 
      &\leq e^{C_{k_{\text{e}}, k_{\text{l}}}^{(\tau_3)}} \left(C_{k_{\text{e}}, k_{\text{l}}} - C_{k_{\text{e}}, k_{\text{l}}}^{(\tau_3)} + 1\right),
     \end{split}
\end{equation}
where $B_{k_{\text{l}}}^{(\tau_3)}$ and $C_{k_{\text{e}}, k_{\text{l}}}^{(\tau_3)}$ are the given local points of $B_{k_{\text{l}}}$ and $C_{k_{\text{e}}, k_{\text{l}}}$, respectively, in the $\tau_3$-th iteration of SCA. As a result, the optimization problem~\eqref{eq: optimization problem W-2} can be reformulated as 
\begin{subequations}
\label{eq: optimization problem W-3}
\begin{equation}
\label{eq: optimal objective function W-3}
    \max_{\left\{\mathbf{W}, A_{k_{\text{l}}}, B_{k_{\text{l}}}, C_{k_{\text{e}}, k_{\text{l}}}, D_{k_{\text{e}}, k_{\text{l}}} \right\}} \sum_{k_{\text{l}}\in\mathcal{K}_{\text{l}}} \tilde{R}^s_{k_{\text{l}}}
\end{equation}
\begin{equation}
\label{eq: optimization before constraint W-3}
    {\rm{s.t.}} \ \  \eqref{eq: W positive semidefinite matrix}, \eqref{eq: active beamforming constraint at BS W-1}, \eqref{eq: legitimate rate numerator W}, \eqref{eq: eve rate numerator W}, \eqref{eq: legitimate user rate constraint W-2}, \eqref{eq: eve rate constraint W-2}, \eqref{eq: legitimate rate denominator constraint Taylor approximations}, \eqref{eq: eve rate denominator constraint Taylor approximations}. \notag
\end{equation}
\end{subequations}

To address the non-convexity of the rank-one constraint~\eqref{eq: W rank one constraint}, the semi-definite relaxation (SDR) technique is employed, which temporarily relaxes the rank-one constraint by ignoring~\eqref{eq: W rank one constraint}, thereby transforming the original problem into a convex SDP problem. {The following theorem demonstrates the equivalence between the problem~\eqref{eq: optimization problem W-3} and the relaxed version.} 

\begin{theorem}
    The rank size of the solution $\mathbf{W_{k_{\text{l}}}}$ of the relaxed version of problem~\eqref{eq: optimization problem W-3}, obtained by removing the rank-one constraint~\eqref{eq: W rank one constraint}, always depends on the rank size of $\mathbf{\Phi}_{\varrho}$, which is expressed as follows:
\end{theorem}
\vspace{-0.6cm}
    \begin{equation}
        \text{rank}\left( \mathbf{W_{k_{\text{l}}}} \right) \leq \text{rank}\left( \mathbf{\Phi}_{\varrho}\right) = 1, \forall k_{\text{l}} \in \mathcal{K}_{\text{l}}.
    \end{equation}
\begin{proof}
    The readers are referred to~\cite[Appendix A]{zhang2023security}, and the details are omitted here.
\end{proof}
Being a convex problem, the relaxed version of the optimization problem~\eqref{eq: optimization problem W-3} can be efficiently addressed using optimization solvers, like CVX. After solving the optimization problem~\eqref{eq: optimization problem W-3}, the optimal active beamforming vectors can be recovered through Cholesky decomposition as $\mathbf{W}^*_{k_{\text{l}}} = \mathbf{w}^*_{k_{\text{l}}}(\mathbf{w}^*_{k_{\text{l}}})^H$. Building on the preceding discussion, the detailed steps of the proposed algorithm are provided in~\textbf{Algorithm~\ref{alg: BS's Beamforming Vectors}}.

\begin{algorithm}
	\renewcommand{\algorithmicrequire}{\textbf{Input:}}
	\renewcommand{\algorithmicensure}{\textbf{Output:}}
	\caption{Proposed SCA-based algorithm for solving BE active beamforming subproblem}
	\label{alg: BS's Beamforming Vectors}
	\begin{algorithmic}[1]
		\State Initialize feasible points $\left\{\mathbf{W}_{k_{\text{l}}}^{(0)}, A_{k_{\text{l}}}^{(0)}, B_{k_{\text{l}}}^{(0)}, C_{k_{\text{e}}, k_{\text{l}}}^{(0)}, D_{k_{\text{e}}, k_{\text{l}}}^{(0)} \right\}$;
            \State Set $\tau_3=0$;
                \Repeat
		      \State Solving problem \eqref{eq: optimization problem W-3} to obtain $\left\{\mathbf{W}_{k_{\text{l}}}^{(\tau_3)}, A_{k_{\text{l}}}^{(\tau_3)}, B_{k_{\text{l}}}^{(\tau_3)}, C_{k_{\text{e}}, k_{\text{l}}}^{(\tau_3)}, D_{k_{\text{e}}, k_{\text{l}}}^{(\tau_3)} \right\}$;
            \State $\tau_3=\tau_3+1$;
		\Until the fractional increase of the objective value is below a predefined threshold $\varepsilon_3$ \textbf{or} $\tau_3 \geq \tau_3^{\text{max}}$.
        \State \Return $\mathbf{W}_{k_{\text{l}}}^*$.
	\end{algorithmic}  
\end{algorithm}

\subsection{Overall AO algorithm design and property analysis}
Based on the algorithms described above, we present the AO algorithm for solving the original problem~\eqref{eq: optimization problem} in~\textbf{Algorithm~\ref{alg: AO}}. Specifically, the MEs positions matrix $\mathbf{R}$, the STARS passive beamforming $\boldsymbol{\Theta}_{\varrho}$, and the BS active beamforming $\mathbf{W}$ are iteratively optimized by invoking \textbf{Algorithm~\ref{alg1: MEs positioning}}, \textbf{Algorithm~\ref{alg1: STARS's Beamforming Vectors}}, and \textbf{Algorithm~\ref{alg: BS's Beamforming Vectors}}, respectively. The iterations terminate once the fractional increase of the sum secrecy rate becomes below a predefined threshold $\varepsilon_0$ or when the maximum number of AO iterations has been reached.  
Next, we analyze the properties of~\textbf{Algorithm~\ref{alg: AO}} in terms of the convergence and the complexity.

\begin{algorithm}
    \renewcommand{\algorithmicrequire}{\textbf{Input:}}
    \renewcommand{\algorithmicensure}{\textbf{Output:}}
    \caption{Proposed AO-based iterative algorithm for solving sum secrecy rate optimization problem}
    \label{alg: AO}
    \begin{algorithmic}[1]
        \State Initialize feasible points $\mathbf{R}^{(0)}, \boldsymbol{\Theta}_{\varrho}^{(0)}, \mathbf{W}^{(0)}$;
        \State Initialize threshold value $\varepsilon_0$;
        \State Set $\tau = 0$;
            \Repeat
            \State Update STARS MEs position matrix $\mathbf{R}$ via \textbf{Algorithm~\ref{alg1: MEs positioning}};
            \State Update STARS passive beamforming matrices $\boldsymbol{\Theta}_{\varrho}$ via \textbf{Algorithm~\ref{alg1: STARS's Beamforming Vectors}};
            \State Update the BS active beamforming matrix $\mathbf{W}$ via \textbf{Algorithm~\ref{alg: BS's Beamforming Vectors}};
            \State Update $\tau = \tau + 1$;
        \Until the fractional increase of the objective value is below $\varepsilon_0$ \textbf{or} $\tau \geq \tau_{\text{max}}$;
        \State \Return $\mathbf{R}^*, \boldsymbol{\Theta}_{\varrho}^*, \mathbf{W}^*$.
    \end{algorithmic}  
\end{algorithm}

\subsubsection{Convergence analysis} 
For prove the convergence of~\textbf{Algorithm~\ref{alg: AO}}, we define  $R_{\text{sum}}\left(\mathbf{R}^{(\tau)},\boldsymbol{\Theta}_{\varrho}^{(\tau)}, \mathbf{W}^{(\tau)}\right)$ as the value of the objective function for the optimization problem~\eqref{eq: optimization problem} at the $\tau$-th AO iteration. The following inequalities hold in each AO iteration $\tau$:
\begin{equation}
\label{eq: achievable rate convergence analysis}
   \begin{split}
    &R_{\text{sum}}\left(\mathbf{R}^{(\tau)},\boldsymbol{\Theta}_{\varrho}^{(\tau)}, \mathbf{W}^{(\tau)}\right)\\
    &\overset{(a)}{\operatorname*{\leq}} R_{\text{sum}}\left(\mathbf{R}^{(\tau+1)},\boldsymbol{\Theta}_{\varrho}^{(\tau)}, \mathbf{W}^{(\tau)}\right)\\
    &\overset{(b)}{\operatorname*{\leq}} R_{\text{sum}}\left(\mathbf{R}^{(\tau+1)},\boldsymbol{\Theta}_{\varrho}^{(\tau+1)}, \mathbf{W}^{(\tau)}\right)\\
    &\overset{(c)}{\operatorname*{\leq}} R_{\text{sum}}\left(\mathbf{R}^{(\tau+1)},\boldsymbol{\Theta}_{\varrho}^{(\tau+1)}, \mathbf{W}^{(\tau+1)}\right).\\
   \end{split}
\end{equation}
Here, (a) holds because $\mathbf{R}$ is updated in the gradient ascent direction of $R_{\text{sum}}$ within the $\tau$-th AO iteration; (b) follows from the fact that the first-order Taylor expansion in~\eqref{eq: legitimate rate denominator constraint first-order Taylor} and~\eqref{eq: eve rate numerator constraint first-order Taylor} are tight at each given local point; (c) is obtained since the objective function in problem~\eqref{eq: optimization problem W-3} is always the lower-bound of the objective function in problem~\eqref{eq: optimization problem W-1}. Overall, since the sum secrecy rate is upper bounded given the maximum transmit power constraint, \textbf{Algorithm~\ref{alg: AO}} is guaranteed to converge to a stable solution of~\eqref{eq: optimization problem} within limited number of iterations.

\subsubsection{Complexity analysis} 
{The computational complexity of \textbf{Algorithm~\ref{alg1: MEs positioning}} is mainly influenced by two critical components, i.e., the gradient calculation of the objective function~\eqref{eq: secrecy rate r-4} and the the process of backtracking line search. The complexity of calculating the gradient $\nabla G\left(\mathbf{R}\right)$ is given by $\mathcal{O} \left(2 N^2 L_{\text{BS}} L_{\text{S},k}\right)$. In addition, the backtracking line search has the complexity of $\mathcal{O} \left(T_{\text{se}}MN \right)$, where $T_{\text{se}}$ represents the number of searching steps. As a result, the overall computational complexity of~\textbf{Algorithm~\ref{alg1: MEs positioning}} is $\mathcal{O}\left(T_1^{\text{o}} T_1^{\text{i}} \left( 2 N^2 L_{\text{BS}} L_{\text{S},k} + T_{\text{se}}MN\right) \right)$, where $T_1^{\text{o}}$ and $T_1^{\text{i}}$ denote the number of outer and inner iterations required for convergence, respectively.} 
For~\textbf{Algorithm~\ref{alg1: STARS's Beamforming Vectors}}, the complexity of the SCA algorithm is given by $\mathcal{O}\left(2T_2^{\text{o}} T_2^{\text{i}} N^{3.5}\right)$~\cite{dinh2010local}, where $T_2^{\text{o}}$ and $T_2^{\text{i}}$ is the number of outer and inner iterations, respectively. Similarly, \textbf{Algorithm~\ref{alg: BS's Beamforming Vectors}} exhibits the complexity of $\mathcal{O}\left(2T_3 M^{3.5}\right)$, where $T_3$ is the number of iterations required for convergence. As such, the overall complexity of the AO algorithm can be summarized as 
$\mathcal{O}\big(T_{\text{AO}} \big(T_1^{\text{o}} T_1^{\text{i}} \big(2 N^2 L_{\text{BS}} L_{\text{S},k} + T_{se}MN\big)+ 2 T_2^{\text{o}} T_2^{\text{i}} N^{3.5} + 2 T_3 M^{3.5}\big)\big)$,
where $T_{\text{AO}}$ is the total number of AO iterations.

\section{Numerical Results}
In this section, we provide the simulation results to evaluate the performance of the proposed ME-STARS-aided secure communication system. 
As illustrated by the $x_s-y_s-z_s$ coordinate in Fig.~\ref{fig: The far-field-response channel model}, the BS is positioned at $(-50, -5, 10)$~meters, with the legitimate users and eavesdroppers randomly scattered within a square region centered at $(0,0,0)$~meters, each side measuring $35$~meters. Additionally, the STARS MEs are capable of moving within a square region with an edge length of $A$ meters. We utilize the geometric channel model, assuming that the number of propagation paths for all channels is uniform, $L_{\text{BS}} = L_{\text{S},k} = L$. Each diagonal element of the path response matrix $\boldsymbol{\Sigma}$ follows CSCG distribution. Specifically, the first diagonal element $\boldsymbol{\Sigma}[1, 1]$ is modeled as $\mathcal{CN}\left(0, \beta_0d_{\text{b}}^{-\alpha_o}\right)$, representing the line-of-sight (LoS) component, where $\kappa_0$ represents the ratio of the average power of the LoS path relative to that of the non-line-of-sight (NLoS) paths, $\beta_o$ represents the path loss at the reference of $1$ meters, $d_{\text{b}}$ denotes the distance between the STARS and the BS, and $\alpha_0$ is the path loss exponent. The diagonal element $\boldsymbol{\Sigma}[p, p]$ with $p = 2, 3, \dots, L$, is modeled as $\mathcal{CN}\left(0, \beta_0d_{\text{s}, k}^{-\alpha_o}/L\right)$, representing the NLoS components, where $d_{\text{s}, k}$ denotes the distance between the STARS and legitimate users/eavesdroppers. Assumptions for the AoDs and AoAs include that they are independent and identically distributed (i.i.d.) random variables, uniformly distributed over the interval $[-\pi/2, \pi/2]$. To prevent coupling effects between the MEs,  we enforce a minimum separation distance of $D_0 = \lambda/2$ between any two MEs, where $\lambda$ is the wavelength of the transmitted signal. Unless otherwise specified,  the simulation parameters used are listed in TABLE~\ref{tab:simulation_parameters}.

\begin{table}[h!]
\caption{Simulation Parameters}
\centering
\label{tab:simulation_parameters}
\begin{tabular}{|p{1cm}|p{5cm}|p{1.1cm}|}
\hline
Parameter & Description & Value \\ \hline
\( \lambda \) & Wavelength & $0.1$~meter \\ \hline
\( L \) & Number of channel paths  & $2$ \\ \hline
\( M \) & Number of BS FPAs & $8$ \\ \hline
\( N \) & Number of STARS MEs & $12$ \\ \hline
\( J_l \) & Number of legitimate users & $2$ \\ \hline
\( J_e \) & Number of eavesdroppers & $2$ \\ \hline
\( \sigma_{\text{l}}^2, \sigma_{\text{e}}^2 \) & Legitimate user/eavesdropper noise power & $-90$~dBm \\ \hline
\( \alpha_0 \) & Path loss exponent & $2.2$ \\ \hline
\( \beta_0 \) & Power gain of the channel at $1$~meter & $-30$~dB \\ \hline
\( \kappa_0 \) & Rician factor & $5$~dB \\ \hline
\( P_{\max} \) & BS maximum transmit power & $30$~dBm \\ \hline
\( A \) & Edge lengths of the movement region $\mathcal{C}$ & \( 2.5\lambda \) \\ \hline
\( \alpha \) & Initial step size for gradient-ascent & 10 \\ \hline
\( \omega_\alpha \) & Step size shrinking parameter & $0.9$ \\ \hline
\( \mu \) & Initial smoothing factor & $1$ \\ \hline
\( \omega_\mu \) & Smoothing factor shrinking parameter & $0.1$ \\ \hline
\( \rho, \eta \) & Initial penalty factor  & \(10^{-6}\) \\ \hline
\( \omega_\rho \) & Scaling factor for $\rho$ & $10$ \\ \hline
\( \omega_\eta \) & Scaling factor for $\eta$ & $10$ \\ \hline
\end{tabular}
\end{table}

To verify the effectiveness of the proposed ME-STARS-aided secure communications scheme, we compare with the following baseline schemes. 
\begin{itemize}
    \item \textbf{Baseline scheme 1 (FPE-STARS):} In this case, the STARS is equipped with $N$ FPEs uniformly spaced with the distance of $\lambda/2$. The active beamforming and passive beamforming are iteratively optimized by employing the \textbf{Algorithm~\ref{alg1: STARS's Beamforming Vectors}} and \textbf{\ref{alg: BS's Beamforming Vectors}}, respectively. 
    \item \textbf{Baseline scheme 2 (ME-RIS):} In this case, two separate RISs, i.e., one dedicated to the transmission (transmission-only RIS) and the other to the reflection (reflection-only RIS), are deployed to provide the full-space coverage. For the sake of a fair comparison, each RIS is assumed to be equipped with $N/2$ MEs, assuming $N$ is an even number. This setup can be viewed as a simplified version of the STARS, where half of the elements operate in transmission mode and the other half in reflection mode.   
    \item \textbf{Baseline scheme 4 (Random position element (RPE)-STARS):}  In this case, we assume that the STARS is equipped with $N$ elements randomly distributed within a movement region of $A\times A$. The passive and active beamforming are iteratively optimized by involving \textbf{Algorithm~\ref{alg1: STARS's Beamforming Vectors}} and \textbf{\ref{alg: BS's Beamforming Vectors}}, respectively.
\end{itemize}

\begin{figure}[htbp]
    \centering
    \begin{subfigure}[b]{0.50\textwidth}
        \centering
        \includegraphics[width=\textwidth]{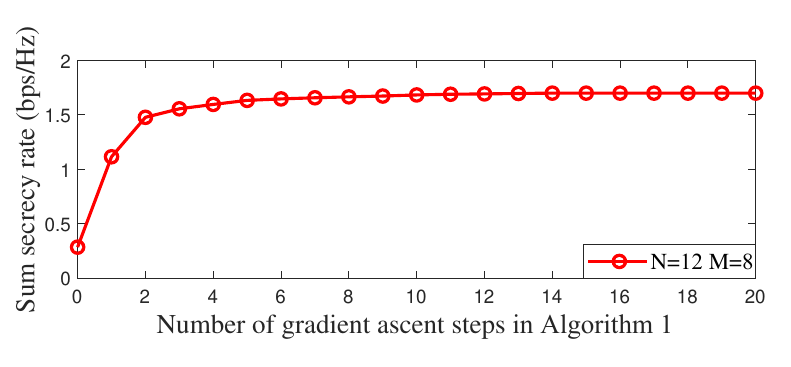}
        \caption{Algorithm 1.}
        \label{fig: Number of inner iterations}
    \end{subfigure}
    \hfill
    \begin{subfigure}[b]{0.50\textwidth}
        \centering
        \includegraphics[width=\textwidth]{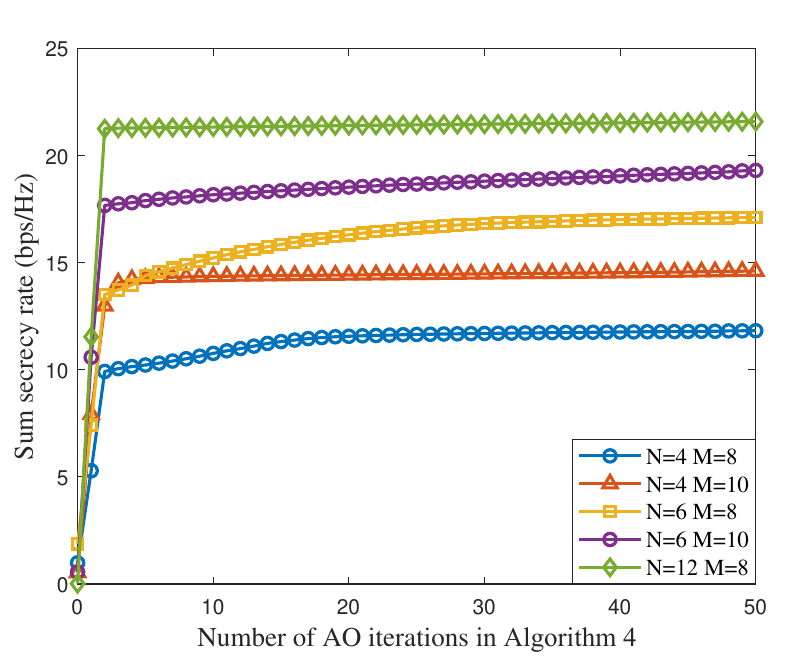}
        \caption{Algorithm 4.}
        \label{fig: Number of outer iterations}
    \end{subfigure}
    \caption{Convergence behavior of the proposed algorithms.}
    \label{fig: The convergence behavior of outer and inner layer}
\end{figure}

Fig.~\ref{fig: Number of inner iterations} illustrates the secrecy rate versus the number of gradient ascent steps in the inner loop of \textbf{Algorithm~\ref{alg1: MEs positioning}}, when $N=12$ and $M=8$. We can find that the secrecy rate keeps increasing after each gradient ascent step and converges within around $10$ iterations. Moreover, Fig.~\ref{fig: Number of outer iterations} demonstrates the secrecy rate versus the the number of AO iterations for different values of $N$ and $M$. It is observed that, the sum secrecy rate consistently increases and stabilizes after approximately $4$ iterations for all the considered $M$ and $N$ setup. It is also evident to see that, for a fixed $M$, increasing $N$ leads to a higher secrecy rate, as more MEs can provide additional DoFs for reconfiguring the channel characteristic and enhancing the passive beamforming gain.

\begin{figure}[h]
    \centering
    \includegraphics[scale=0.68]{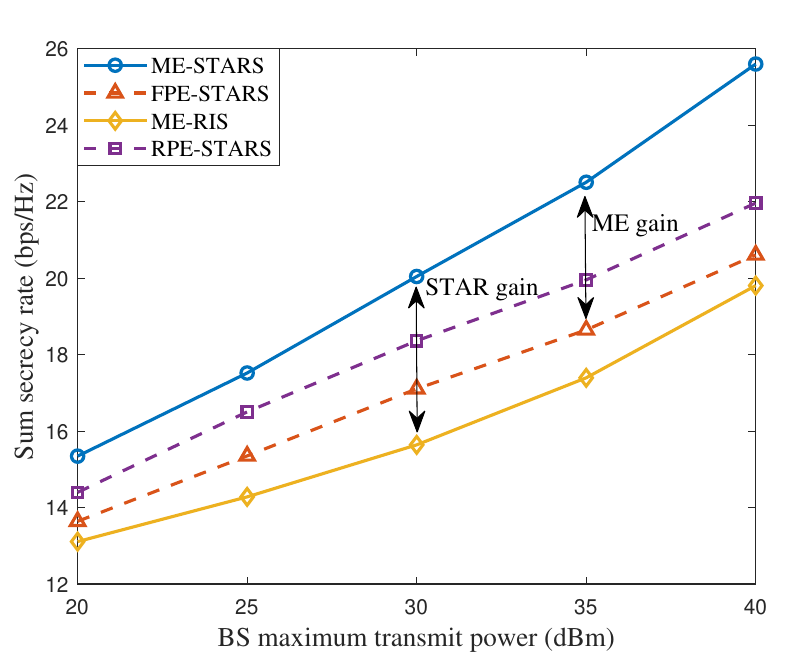}
    \caption{Sum secrecy rate versus BS maximum transmit power.}
    \label{fig: Sum secrecy rate versus BS maximum transmit power}
\end{figure}

In Fig.~\ref{fig: Sum secrecy rate versus BS maximum transmit power}, we compare the achievable sum secrecy rate of different schemes versus the BS maximum transmit power $P_{\text{max}}$. It can be first observed that, the sum secrecy rates of all schemes exhibit a monotonically increasing trend with the increment of $P_{\text{max}}$. We can also find that the proposed ME-STARS scheme significantly outperforms the FPE-STARS counterpart by around $25\%$, which is consistent with our previous analysis that the movable positions of STARS elements can bring in higher spatial DoFs. Moreover, the ME-STARS also demonstrates appealing performance gain compared to the ME-RIS benchmark. The reason behind this is that, each element of the ME-RISs can only work in the transmission or the reflection mode, which restricts the flexibility for simultaneously adjusting the transmission and reflection coefficients as performed by the STARS. One can also observe from Fig.~\ref{fig: Sum secrecy rate versus BS maximum transmit power} that, the ME-STARS always outperforms the RPE-STARS under all BS transmit power, which confirms the effectiveness of \textbf{Algorithm~1} for optimizing MEs positions.

\begin{figure}
    \centering
    \includegraphics[scale=0.68]{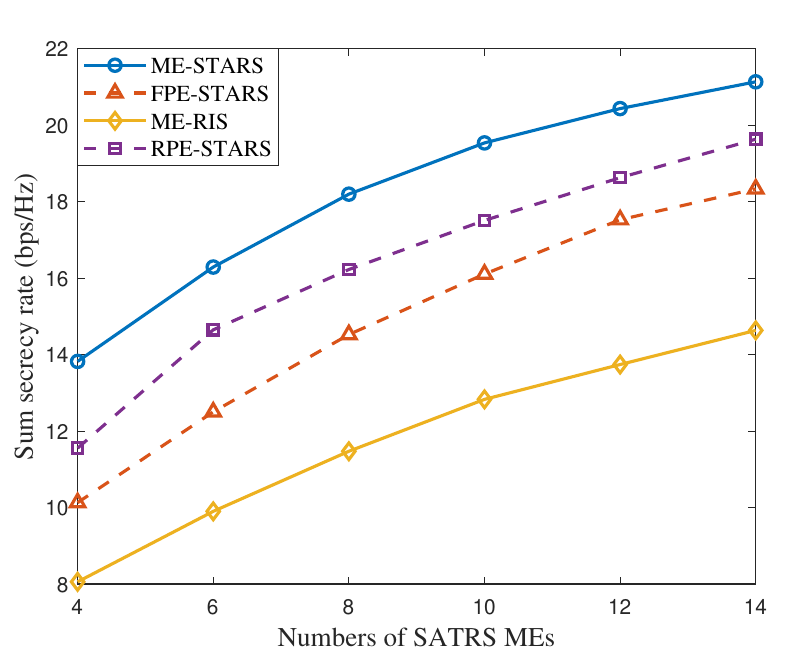}
    \caption{Sum secrecy rate versus number of STARS MEs.}
    \label{fig: Sum secrecy rate versus number of STARS MEs}
\end{figure}

In Fig.~\ref{fig: Sum secrecy rate versus number of STARS MEs}, we investigate the achievable sum secrecy rate of different schemes versus the number of STARS MEs $N$. As $N$ increases, all schemes show a increasing trend in terms of the sum secrecy rate. This is due to the fact that, the larger number of MEs lead to higher passive beamforming gain, and thus simultaneously improve the data rate to the legitimate users and reduce the information leakage to the eavesdroppers. It is also observed that, the increasing rate of the sum secrecy rate achieved by the ME-STARS gets smaller with the increment of $N$. This can be explained as follows. Since the MEs movable region is restricted, the number of local optimal positions for maximizing the sum secrecy rate is limited. When $N$ becomes larger than the number of local maxima, not all MEs can be moved to the preferable positions, which leads to the performance saturation.

\begin{figure}
    \centering
    \includegraphics[scale=0.68]{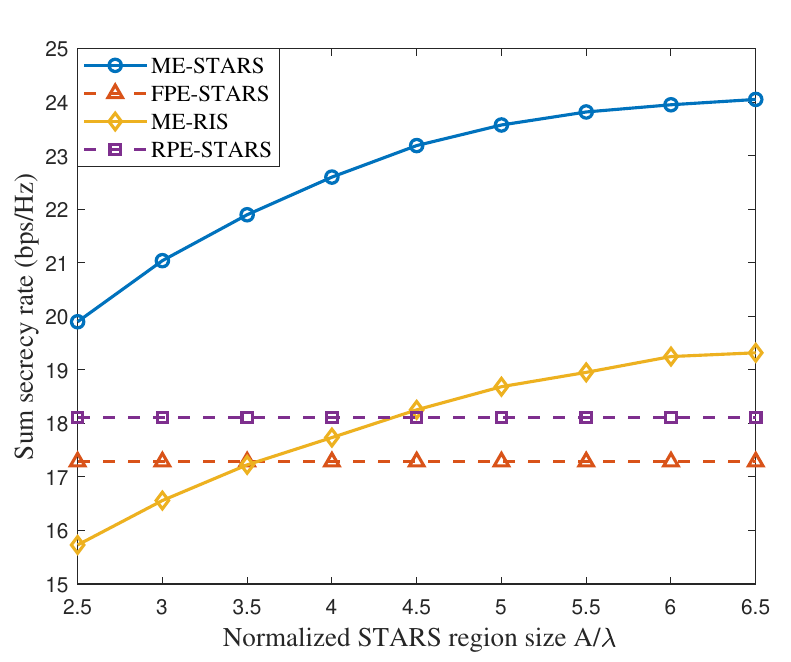}
    \caption{Sum secrecy rate versus normalized STARS region size.}
    \label{fig: Sum secrecy rate versus number of STARS MEs region size}
\end{figure}

\begin{figure}
    \centering
    \includegraphics[scale=0.68]{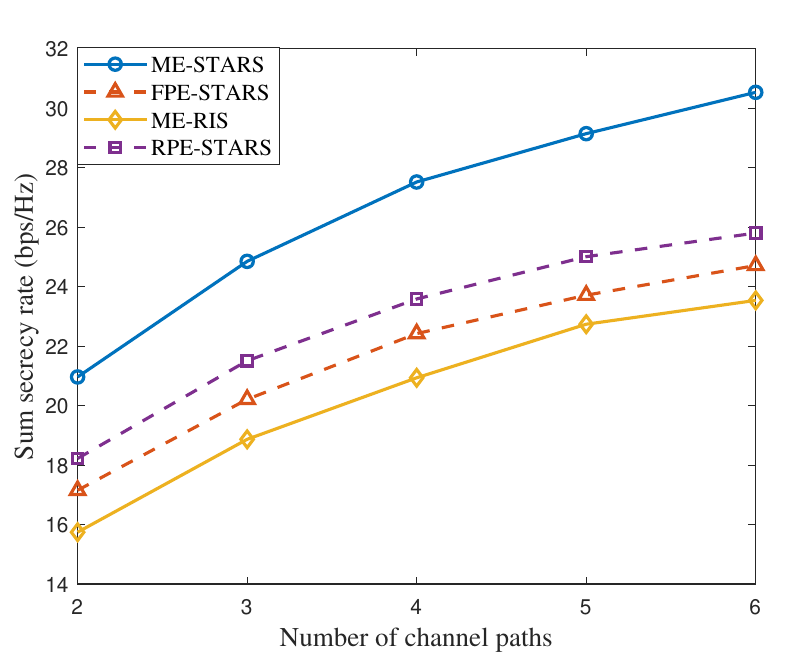}
    \caption{Sum secrecy rate versus number of channel paths.}
    \label{fig: Sum secrecy rate versus number of channel paths}
\end{figure}

In Fig.~\ref{fig: Sum secrecy rate versus number of STARS MEs region size}, we study the achievable sum secrecy rate of different schemes versus the normalized ME-STARS movable region size $A/\lambda$. It is observed that the sum secrecy rate for all the ME schemes improves with the expanded movable region size. This is because, a larger movable region can provide higher spatial diversity and enable MEs to adjust their positions more flexibly for constructing preferable channel conditions. It is also worthy note that the performance gain gets limited when the region size gets sufficiently large. This is because, when the number of optimal positions for the given number of MEs is adequate, enlarging the region size will not bring in any further performance gain. This implies the importance for properly designing the ME-STARS aperture size in practical implementations.

In Fig.~\ref{fig: Sum secrecy rate versus number of channel paths}, we demonstrate the sum secrecy rate achieved by different schemes versus the number of channel paths $L$. It is first seen that all the considered schemes reap obvious secrecy rate improvement with the enlarged $L$. This can be explained as follows. With the increment of $L$, the correlations of legitimate users/eavesdroppers channels get lower. As such, the multi-user interference as well as the information leakage are reduced by exploiting the beamforming gain, which leads to higher sum secrecy rate. It is also observed that the ME-STARS shows the most pronounced secrecy rate improvement with larger $L$. This is because, larger $L$ also leads to the increased number of local optima of MEs positions within the given limited region, thanks to the enhanced small-scale fading.

\section{Conclusion}
In this paper, we investigated a novel ME-STARS-aided secure communication system, where MEs were deployed at the STARS for enhancing the PLS by exploiting the extra spatial DoFs. Against the full-space eavesdropping, we maximized the sum secrecy rate of legitimate users via the joint optimization of the MEs positions, as well as the active and passive beamforming. An AO-based iterative approach was invoked for solving the resultant non-convex optimization problem involving highly-coupled variables. By decomposing the original problem into three subproblems, the gradient ascent algorithm was developed for addressing the MEs positions optimization, while the SCA technique was employed for solving the active and passive beamforming subproblems. Numerical results demonstrated that ME-STARS significantly outperformes the conventional FPE-STARS and the ME-RIS in terms of the secrecy performance. Moreover, the sum secrecy rate achieved by the ME-STARS was demonstrated to increase with enlarged movable region size and saturate within a restricted upper bound. 

\begin{appendices} 
\section{Derivation of $\nabla G\left(\mathbf{R}\right)$}
\setcounter{equation}{0}
\renewcommand\theequation{A.\arabic{equation}}
\label{app: A}

This section provides a detailed derivation of the gradients used in the optimization process, focusing on matrix operations. The objective function's gradient, denoted by $\nabla G\left(\mathbf{R}\right)$, can be expressed as 
\begin{equation}
\label{eq: The gradient of the objective function}
    \nabla G\left(\mathbf{R}\right) = \left[\nabla_{\mathbf{r}_1} G\left(\mathbf{R}\right), \nabla_{\mathbf{r}_2} G\left(\mathbf{R}\right), \ldots, \nabla_{\mathbf{r}_N} G\left(\mathbf{R}\right) \right],
\end{equation}
where each column $\nabla_{\mathbf{r}_n} G\left(\mathbf{R}\right)$ represents the gradient w.r.t. parameter $\mathbf{r}_n$. The specific expression for $\nabla_{\mathbf{r}_n} G\left(\mathbf{R}\right)$ is given by
\begin{equation}
\label{eq: the n-th column of objective function}
    \frac{\partial G\left({\mathbf{R}}\right)}{\partial \mathbf{r}_n} = \sum_{k_{\text{l}}\in\mathcal{K}_{\text{l}}} \frac{\partial R^{\text{s}}_{k_{\text{l}}}}{\partial \mathbf{r}_n} - \rho\sum_{i\in\mathcal{I}} \lambda_{i}\frac{\partial f_{i}(\mathbf{R})}{\partial \mathbf{r}_n},
\end{equation}
where
\begin{equation}
\label{eq: the n-th column of objective function}
    \frac{\partial R^{\text{s}}_{k_{\text{l}}}}{\partial \mathbf{r}_n} = \frac{\partial R_{k_{\text{l}}}}{\partial \mathbf{r}_n} - \frac{\partial R_{e^{\text{max}}, k_{\text{l}}}^{\text{e}}}{\partial \mathbf{r}_n}, 
\end{equation}

\begin{equation}
\label{eq: the deviation of minimum MEs distance}
    \frac{\partial f_{n,n'}\left(\mathbf{R}\right)}{\partial \mathbf{r}_n} =  - \frac{\mathbf{r}_n - \mathbf{r}_{n'}}{\sqrt{(\mathbf{r}_n - \mathbf{r}_{n'})^H(\mathbf{r}_n - \mathbf{r}_{n'})}},
\end{equation}

\begin{equation}
\label{eq: the deviation of rate threshold}
    \frac{\partial f_{k_{\text{l}}}\left(\mathbf{R}\right)}{\partial \mathbf{r}_n} = - \frac{\partial R_{k_{\text{l}}}}{\partial \mathbf{r}_n}, ~~\frac{\partial f_{k_{\text{e}}, k_{\text{l}}}\left(\mathbf{R}\right)}{\partial \mathbf{r}_n} =  \frac{\partial R_{k_{\text{e}}, k_{\text{l}}}^{\text{e}}}{\partial \mathbf{r}_n},
\end{equation}
and
\begin{equation}
\label{eq: the n-th column of objective function}
    \lambda_{i} = \frac{e^{f_i(\mathbf{R})/\mu}}{1 + e^{f_i(\mathbf{R})/\mu}}.
\end{equation}

Using the chain rule, the gradient of legitimate user $k_{\text{l}}$'s data rate can be expressed as
\begin{equation}
\label{eq: the deviation of legitimate rate}
    \frac{\partial R_{k_{\text{l}}}}{\partial \mathbf{r}_n} = \frac{\frac{\partial a_{k_{\text{l}}, k_{\text{l}}}}{\partial \mathbf{r}_n} \left(a_{k_{\text{l}}, \bar{k}_{\text{l}}} + \sigma_{\text{l}}\right) - a_{k_{\text{l}}, k_{\text{l}}}\frac{\partial a_{k_{\text{l}}, \bar{k}_{\text{l}}}}{\partial \mathbf{r}_n}}{\ln 2 \left(a_{k_{\text{l}}, k_{\text{l}}} + a_{k_{\text{l}}, \bar{k}_{\text{l}}} + \sigma_{\text{l}}\right)\left(a_{k_{\text{l}}, \bar{k}_{\text{l}}} + \sigma_{\text{l}}\right)},
\end{equation}
where $a_{k_{\text{l}}, k_{\text{l}}} = \left|\mathbf{q}_{\varrho}^H\mathbf{V}_{k_{\text{l}}}(\mathbf{r})\mathbf{w}_{k_{\text{l}}}\right|^2$. For the derivation of the gradient vector of $a_{k_{\text{l}}, k_{\text{l}}}$ at the point $\mathbf{r}_n$, i.e. $\nabla_{\mathbf{r}_n}a_{k_{\text{l}}, k_{\text{l}}} = \left[\frac{\partial a_{k_{\text{e}}, k_{\text{l}}}}{\partial x_n}  \frac{\partial a_{k_{\text{e}}, k_{\text{l}}}}{\partial y_n}  \right]$, we first define constant $z_{n, k_{\text{l}}}$ as
\begin{equation}
\label{eq: constant z of legitimate}
    z_{n, k_{\text{l}}} = \sum_{n' \neq n}^N \mathbf{x}_{n',k_{\text{l}}}\mathbf{F}_{k_{\text{l}}}(\mathbf{r}_{n'})\mathbf{y}_{k_{\text{l}}} = \vert z_{n, k_{\text{l}}} \vert e^{\angle z_{n, k_{\text{l}}}},
\end{equation}
where $\vert z_{n, k_{\text{l}}} \vert$ represents the amplitude of $z_{n, k_{\text{l}}}$, and $\angle z_{n, k_{\text{l}}}$ represents the phase of $z_{n, k_{\text{l}}}$. Similarly, we define $x_{n, k_{\text{l}}}^p = \vert x_{n, k_{\text{l}}}^p \vert e^{\angle x_{n, k_{\text{l}}}^p}$ as the $p$-th element of $x_{n, k_{\text{l}}}$ with the amplitude $\vert x_{n, k_{\text{l}}}^p \vert$ and the phase $\angle x_{n, k_{\text{l}}}^p$, and $y_{k_{\text{l}}}^o = \vert y_{k_{\text{l}}}^o \vert e^{\angle y_{k_{\text{l}}}^o}$ as the $o$-th element of $y_{n, k_{\text{l}}}$ with the amplitude $\vert y_{k_{\text{l}}}^o \vert$ and the phase $\angle y_{k_{\text{l}}}^o$. Thus, $a_{k_{\text{l}}, k_{\text{l}}}$ can be written as 
\begin{equation}
\label{eq: useful power}
    a_{k_{\text{l}}, k_{\text{l}}} = \left\vert \sum_{o=1}^{L_{\text{BS}}} \sum_{p=1}^{L_{\text{S}, k}} \vert x_{n, k_{\text{l}}}^p \vert \vert y_{k_{\text{l}}}^o \vert e^{j\psi_{k_{\text{l}}}(\mathbf{r}_n)} + \vert z_{n, k_{\text{l}}} \vert e^{\angle z_{n, k_{\text{l}}}}\right\vert^2,
\end{equation}
where 
\begin{equation}
\label{eq: angle of useful power}
    \psi_{k_{\text{l}}, k_{\text{l}}}^{p, o}(\mathbf{r}_n) = \frac{2\pi}{\lambda} \left(\rho_{\text{S}, k_{\text{l}}}^{p}(\mathbf{r}_n) - \rho_{\text{S}, \text{in}}^{o}(\mathbf{r}_n)\right) + \angle x_{n, k_{\text{l}}}^p + \angle y_{k_{\text{l}}}^o.
\end{equation}
Then, the partial deviation of $a_{k_{\text{l}}, k_{\text{l}}}$ w.r.t. $x_n$ can be expressed as
\begin{equation}
\label{eq: the partial deviation of a w.r.t. x}
    \frac{\partial a_{k_{\text{l}}, k_{\text{l}}}}{\partial x_n} = 2\vert z_{n, k_{\text{l}}} \vert \sum_{o=1}^{L_{\text{BS}}} \sum_{p=1}^{L_{\text{S}, k}} \frac{\partial \psi^{p, o}_{k_{\text{l}}, k_{\text{l}}}(\mathbf{r}_n)}{\partial x_n} \vert x_{n, k_{\text{l}}}^p \vert \vert y_{k_{\text{l}}}^o \vert \xi^{p, o}_{k_{\text{l}}, k_{\text{l}}}(\mathbf{r}_n),
\end{equation}
where $\xi_{k_{\text{l}}, k_{\text{l}}}^{p, o}(\mathbf{r}_n) = \sin{\left(\angle z_{n, k_{\text{l}}} - \psi^{p, o}_{k_{\text{l}}, k_{\text{l}}}(\mathbf{r}_n)\right)}$, $\partial \psi^{p, o}_{k_{\text{l}}, k_{\text{l}}}/\partial x_n = \frac{2\pi}{\lambda} \left(\text{cos}\theta_{\text{S}, k_{\text{l}}}^p\text{sin}\phi_{\text{S}, k_{\text{l}}}^p - \text{cos}\theta_{\text{s,b}}^o\text{sin}\phi_{\text{s,b}}^o\right)$. And the partial deviation of $a_{k_{\text{l}}, k_{\text{l}}}$ w.r.t. $y_n$ can be expressed as
\begin{equation}
\label{eq: the partial deviation of a w.r.t. y}
    \frac{\partial a_{k_{\text{l}}, k_{\text{l}}}}{\partial y_n} = 2\vert z_{n, k_{\text{l}}} \vert \sum_{o=1}^{L_{\text{BS}}} \sum_{p=1}^{L_{\text{S}, k}} \frac{\partial \psi^{p, o}_{k_{\text{l}}, k_{\text{l}}}(\mathbf{r}_n)}{\partial y_n} \vert x_{n, k_{\text{l}}}^p \vert \vert y_{k_{\text{l}}}^o \vert \xi^{p, o}_{k_{\text{l}}, k_{\text{l}}}(\mathbf{r}_n),
\end{equation}
where $\partial \psi^{p, o}_{k_{\text{l}}, k_{\text{l}}}(\mathbf{r}_n)/\partial y_n = \frac{2\pi}{\lambda} \left(\text{sin}\theta_{\text{S}, k_{\text{l}}}^p - \text{sin}\theta_{\text{s.b}}^o\right)$.

The derivation of $\partial R_{k_{\text{e}}, k_{\text{l}}}^{\text{e}}/\partial \mathbf{r}_n$ follows a similar procedure, and the detailed steps are omitted for brevity.
\end{appendices}

\bibliographystyle{IEEEtran}
\bibliography{mybib}
\end{document}